\documentclass{article}
\usepackage[T1]{fontenc}
\usepackage{lmodern, microtype}
\usepackage{geometry}
\usepackage{setspace}
\usepackage{rotating}
\usepackage[ngerman, english]{babel}
\usepackage{natbib}
\usepackage{caption}
\usepackage{tabularx}
\usepackage{floatrow}
\usepackage{soul}
\usepackage{url}
\usepackage{booktabs}
\usepackage{dcolumn}
\usepackage{graphicx}

\newcolumntype{C}{>{\centering\arraybackslash}X}
\newcolumntype{L}{>{\raggedright\arraybackslash}X}

\title{Zombies in the Loop? Humans Trust Untrustworthy AI-Advisors for Ethical Decisions\thanks{Funding by the Bavarian Research Institute for Digital Transformation and the German Federal Ministry of Education and Research (16SV8370) is gratefully acknowledged.}}
\author{
	Sebastian Kr\"{u}gel\thanks{Faculty of Computer Science, \foreignlanguage{ngerman}{Technische Hochschule Ingolstadt} and TUM School of Governance, \foreignlanguage{ngerman}{Technische Universit\"{a}t M\"{u}nchen}, sebastian.kruegel@tum.de.}
	\and Andreas Ostermaier\thanks{Department of Business \& Management, University of Southern Denmark, ostermaier@sam.sdu.dk.}
	\and Matthias Uhl\thanks{Faculty of Computer Science, \foreignlanguage{ngerman}{Technische Hochschule Ingolstadt}, Matthias.Uhl@thi.de.}
}
\date{}

\begin{document}
	\maketitle
	
	\thispagestyle{empty}
	
	\begin{abstract}
		Departing from the claim that AI needs to be trustworthy, we find that ethical advice from an AI-powered algorithm is trusted even when its users know nothing about its training data and when they learn information about it that warrants distrust. We conducted online experiments where the subjects took the role of decision-makers who received advice from an algorithm on how to deal with an ethical dilemma. We manipulated the information about the algorithm and studied its influence. Our findings suggest that AI is overtrusted rather than distrusted. We suggest digital literacy as a potential remedy to ensure the responsible use of AI.
		
		\bigskip
		
	\end{abstract}
	
	\clearpage
	
	\setcounter{page}{1}
	
	\vspace*{12pt}
	
	\begin{center}
		\LARGE Zombies in the Loop? Humans Trust Untrustworthy AI-Advisors for Ethical Decisions
	\end{center}
	
	\doublespacing
	
	\section*{Introduction}
	
	The ethics guidelines and principles for AI issued by government agencies, industry associations, and business companies are unified by the claim that AI should be trustworthy \citep{JobinEtAl2019}. For example, the U.S. government advances the ``development and use of trustworthy AI'' \citep{NAIIO2021}. Likewise, the European Union has developed ``ethics guidelines for trustworthy AI'' \citep{EC2019}. Trustworthiness is considered necessary for AI to earn trust, which in turn ``is needed for its fruitful, pervasive use in our daily lives'' \citep[][p.~2]{IEEE2017}. While this claim is intuitive from a normative viewpoint, it is also based on the empirical hypothesis that trust requires trustworthiness. However, there is little evidence to support this argument for AI. The purpose of this study is to investigate how sensitive users are to the trustworthiness of AI-powered learning moral advisors.
	
	AI-powered algorithms have conquered areas such as personnel recruitment, the allocation of loans, penal sentencing, or autonomous driving \citep{RahwanEtAl2019}. They make and help us make highly consequential decisions and have practically turned into ethical agents \citep{Whitby2011, VoiklisEtAl2016}. In particular, the algorithm can serve as a moral advisor to its human user, who still makes the decision and accounts for it. Human involvement in algorithmic decision-making enhances perceived control over the algorithm and has been found to increase trust \citep{DietvorstEtAl2015, BurtonEtAl2020}. The human in the loop is therefore considered a building block in creating trustworthy AI \citep{EC2019}. However, this argument assumes that the human user does not na\"{i}vely trust the algorithm regardless of how trustworthy it is but carefully checks its advice and makes his or her own decision if a red flag rises.
	
	In the case of learning AI, the transparency and integrity of the training data are minimum requirements for an algorithm to be trustworthy \citep{IEEE2017, LepriEtAl2018, EC2019}. A major concern about algorithms, which transparency can help mitigate, is that they are biased \citep{MittelstadtEtAl2016, JobinEtAl2019}. We explore, first, whether users trust the moral advice of an algorithm if they know nothing about how it generates advice. Our benchmark is AI-generated advice that is based on the judgments of impartial human advisors. While human judgment is notoriously untransparent, the notion of the impartial advisor evokes the ideal observer and gives the advisee an idea of how the advice comes about \citep{Jollimore2021}. Hence, an algorithm is more transparent to its users if they know it was trained on judgments of impartial human advisors than if they know nothing about how it generates moral advice.
	
	Second, we study whether users trust the algorithm's advice when the integrity of its training data is dubious. Specifically, we assume that moral advice from convicted criminals is distrusted by many. Indeed, moral judgment is impaired by pathological traits \citep{CampbellEtAl2009, JonasonEtAl2015, Blair2017}, which are common among criminal offenders. There is also recent evidence that criminal offenders' judgment is biased relative to the average population \citep{KoenigsEtAl2012, YoungEtAl2012, LahatEtAl2015}. Of course, crime does not necessarily arise from a lack of moral judgment. People often know what is right but do the wrong thing nonetheless. We find it still reasonable to assume that training data from convicts are perceived as biased, and discrimination in education, employment, and housing showcases the deep-seated distrust against convicts \citep{SokoloffSchenckFontaine2017, EvansEtAl2019, SugieEtAl2020}.
	
	We report three experiments to examine the acceptance of moral advice from an algorithm. In each experiment, the subjects were advised by an algorithm in choosing between two options in an ethical dilemma. In the first study, the algorithm was described as modeled on the judgments of impartial human advisors, and the subjects followed the algorithm's advice just like advice from an impartial human advisor. In the second study, they were told that it was unkown what the algorithm's advice would be based on, but they followed it indifferently. In the third study, we told them that the algorithm mimicked convicted criminals. The subjects now turned out to make their judgments independently of the human advisor (i.e., the convict who advised them), which confirms our assumption that advice from criminals is distrusted. By contrast, the algorithm said to imitate convicted criminals invariably influenced the advisees even then.
	
	Our results show that users readily accept ethical advice from algorithms even when they know nothing about their training data or when these are presumably biased. This insight challenges the intuition that AI needs to be trustworthy to be trusted. In turn, it corresponds with evidence for algorithm appreciation from outside the moral domain, which suggests that people are more receptive to or, to put it negatively, unreflective of algorithms than one might expect \citep{LoggEtAl2019}. It is noteworthy how insensitive users are to the information provided in the human--machine interaction, which we manipulate. In summary, our findings provide first evidence that algorithms are accepted as moral advisors, on the one hand. On the other hand, they suggest that we think about how to ensure that AI-powered algorithms are used responsibly---e.g., by improving digital literacy. Our study thus contributes to the literature on AI ethics.
	
	\section*{Procedure}
	
	We ran our three experiments on CloudResearch in March 2021. CloudResearch is an online platform to recruit subjects and conduct studies \citep{LitmanEtAl2017}. We recruited a total of 2,017 U.S. residents from CloudResearch's Prime Panels \citep{ChandlerEtAl2019}. Online platforms such as CloudResearch were found to provide reliable and valid results across a wide range of tasks and measures and they have been frequently used in the social sciences \citep{GoodmanEtAl2013, HauserSchwarz2016, ChandlerEtAl2019}. Prime Panels members must pass default screening questions to take part in a study, and they are more diverse and more representative of the U.S. population than MTurk participants in terms of age, family background, religion, education, and political attitudes \citep{ChandlerEtAl2019}. Each experiment took the subjects about five minutes to complete, and they were compensated with a fixed US\$1.25.
	
	We programmed our experiments in Qualtrics. The factorial design resulted in multiple experimental conditions. We used Qualtrics to randomly assign the subjects each to one condition, and CloudResearch to preclude repeated participation. The experiment consisted of four parts. First, we obtained informed consent from the subjects, while we guaranteed confidentiality and voluntary participation (Screens \#1 and \#2). Second, they answered the focal question about an ethical dilemma with the advice of an algorithm or human advisor, which we manipulated (Screen \#3). Third, we posed a question to probe the subjects' understanding and attention (Screen \#4). Fourth, they were asked a series of post-experimental questions about their ethical mindset, their attitude to artificial intelligence, and demographic data (Screens \#5 to \#11). The appendix includes screenshots and further details and technicalities.
	
	We obtained ethical approval for the studies from the institutional review board of the German Association for Experimental Economic Research (\url{https://www.gfew.de}). Each study was pre-registered at AsPredicted (\url{https://www.aspredicted.org}), where we specified the experimental conditions, the key variables, and the planned analyses. Moreover, we determined that the analyses would be restricted to subjects who answered the comprehension question correctly, and we set the number of subjects per condition to fifty. Anticipating that some would fail to prove their comprehension, we recruited more subjects; we attained the planned number of subjects after excluding about twenty percent, who had answered the comprehension question incorrectly. The final sample totaled 1,593 subjects for the three studies. The URL addresses to access the pre-registration documents are included in the appendix.
	
	\section*{Study~1}
	
	Do decision-makers accept ethical advice from a machine? On the one hand, there is plenty of evidence for algorithm aversion. When offered a choice, people would rather take advice from a human than from an artificial advisor, even after seeing the latter outperform the former. They lose confidence in algorithms more easily, more persistently, and more than they reasonably should when these err \citep{DietvorstEtAl2015}. They prefer that human decision-makers rather than machines make moral decisions \citep{BigmanGray2018}, delegate decisions rather to humans, and prefer that others delegate decisions to humans \citep{GogollUhl2018}. Approaches to reduce algorithm aversion include digital literacy, behavioral design, and control by human involvement \citep{DietvorstEtAl2018, BurtonEtAl2020}. The concept of the human in the loop both illustrates skepticism about algorithms and is a prime example of a remedy to mitigate it.
	
	On the other hand, there is also evidence that people trust artificial more than human advice. As a counterpart to \citeauthor{DietvorstEtAl2015}'s \citeyearpar{DietvorstEtAl2015} algorithm aversion, \citet{LoggEtAl2019} coined the term algorithm appreciation. In a series of experiments, they showed that people give more weight to an estimate by an algorithm than by another person in adjusting their own judgments in various domains, ranging from weight estimates to romantic matches, whether they are provided with either one estimate or both. Likewise, news audiences were found to prefer news to be selected for them by algorithms over the selection by editors \citep{ThurmanEtAl2019}. Despite their concerns about algorithms, people often choose automatically taken decisions rather than decisions by human experts or they are indifferent \citep{AraujoEtAl2020}. Moreover, they appear to trust algorithms to be as cooperative as human interaction partners \citep{KarpusEtAl2021}.
	
	If people prefer human over artificial decision-makers in the moral domain \citep{BigmanGray2018, GogollUhl2018}, this does not imply that they will not accept advice from an algorithm when they receive it. \citeauthor{DietvorstEtAl2015}'s \citeyearpar{DietvorstEtAl2015} and \citeauthor{LoggEtAl2019}'s \citeyearpar{LoggEtAl2019} studies resemble ours in that in both a human is advised by a machine. The two studies differ in whether the user has the algorithm seen erring before being advised. \citeauthor{LoggEtAl2019}\ note that advisees follow artificial more than human advice in \citeauthor{DietvorstEtAl2015}'s study until they have seen it erring, and this case is relevant because many decisions are made without feedback on whether the advice was right. Moreover, it is harder to agree intersubjectively on whether moral advice was erroneous than whether a factual prediction was. Neither study considers advice-taking in the moral domain, though, and the question of whether moral advice is accepted from algorithms thus remains open.
	
	To answer this question, imagine a situation where someone needs to make a decision that is consequential for someone else and thus clearly has an ethical dimension (e.g., a recruiter who selects an applicant for a job). In particular, suppose that there are two options to choose from, which leave the decision-maker in a moral dilemma. Moreover, there is an AI-powered algorithm to advise the decision-maker, which either encourages or discourages the choice of one option over the other. If the decision-maker heeds the algorithm's advice, we should see him or her tend more or less to choose one of the two options, depending on the algorithm's advice. Conversely, if the decision-maker is averse to following the algorithm's advice, he or she should disregard it, the decision will be made independently of the advice, and no association between the advice and the inclination to choose the option encouraged by the algorithm should be discerned.
	
	While an association between the algorithm's advice and the human decision would confirm the expectation that moral advice is accepted from an algorithm, it would not show either algorithm aversion or appreciation strictly speaking, which are defined relative to the effect of human advice \citep{DietvorstEtAl2015, LoggEtAl2019}. To see whether the decision-maker heeds the algorithm's advice more or less than human advice, or equally, provided that he or she heeds the advice in the first place, we need to consider the same setting with a human advisor instead. Naturally, human advice that encourages the choice of one option over the other should increase the inclination to choose that option, whereas discouraging advice should reduce it. Algorithm aversion would then result in a weaker effect of the algorithm's relative to the otherwise identical human advice; algorithm appreciation would result in a relatively stronger effect.
	
	\subsection*{Method}
	
	To explore how decision-makers respond to advice by an human-trained AI-powered algorithm, we designed an experiment which required the subjects to make a decision that would benefit either a friend or a stranger. We vignetted three different scenarios, one in the business, health, and legal domain respectively, to preclude that our results would be driven by some specific situation. Each scenario featured the same trade-off between friendship and duty, and each subject was randomly assigned to one of those three scenarios. In the business scenario, for example, the subject was placed in the role of the recruiter of a company. That recruiter had two applicants shortlisted to fill a vacancy: a friend of hers and a stranger. She would then decide whom to hire. There was no further information about the two applicants given, other than that the recruiter found the stranger more eligible, but that she also felt obligated to her friend.
	
	The vignette further introduced an AI-powered algorithm to advise the recruiter. The algorithm was described to the recruiter as imitating human decisions that were based on the moral judgments of impartial human advisors in a situation like hers. Moreover, the recruiter was told that the applicants did not know about the algorithm, and that no one would ever learn whether she followed the algorithm's advice. We thus prevented the recruiter from feeling controlled rather than advised. We manipulated the advice by either saying that it was ethically acceptable or unacceptable for the recruiter to hire her friend. Whether the advice was in favor of or against her friend, was randomly determined by the experimental software for each recruiter. The recruiter was then asked to indicate how much she agreed with the statement that she would hire the stranger, not her friend, on a scale ranging from 0 (``fully disagree'') to 100 (``fully agree'').
	
	To establish that the algorithm's advice influences the recruiter in her decision, it is enough to show that the subjects' decisions differ depending on the algorithm's advice. However, we would naturally expect a similar, potentially larger effect of human advice. For comparison, we provided another set of subjects with the exact same vignette, except that the recruiter was advised by an impartial human advisor rather than by an algorithm whose advice was based on the judgments of impartial human advisors. By this description, we made the algorithm's advice resemble the human advice to isolate the effect of the type of advisor. Like the algorithm, the human advisor would either tell the recruiter that it was acceptable or that it was unacceptable to hire her friend, and no one would ever learn how the recruiter decided. Whether the advice was in favor of or against the recruiter's friend was again randomly determined by the software.
	
	The law and health scenarios were vignetted and administered in the same way as the business scenario, where the subject was put into the shoes of the recruiter. In the law scenario, the subject took the role of a prosecutor responsible for prosecuting money laundering in financial services firms. That prosecutor had to decide which of two suspicious firms to raid---one run by a friend; the other, by a stranger. In the health scenario, the subject was responsible for compiling the list of recipients of kidney donations, and she needed to choose whether to allot the next available position on the list either to a friend of hers or to a stranger. In both cases, we varied the scenario, but kept everything else equal. The screenshots containing the instructions for all three scenarios are reprinted in the appendix. The screens were identical for all possible cases except for Screen \#3, which varied between scenarios and treatment conditions.
	
	\subsection*{Results}
	
	We collected data from a total of 825 subjects. To ensure that the subjects whose answers we were going to analyze had diligently read their vignette and made their decision, we asked a comprehension question to probe their attention and understanding. Our results are based on the 633 subjects who answered this question correctly. 38 percent of these 633 subjects indicated that they were male, and their age averaged 41.7 years, within a range from 18 to 93 years. Our experiment employs a $2 \times 2 \times 3$ factorial design with either a human advisor or the algorithm, who advises that it is acceptable or unacceptable for the decision-maker to favor her friend in one of three scenarios. The subjects were randomly assigned to one of the twelve resulting conditions. Our main interest is in how much the subjects agreed, on a scale ranging from 0 to 100, that they would rather decide in favor of the stranger than the friend in each condition.
	
\begin{figure}[t]
		\ffigbox[\FBwidth]{
			\caption{Inclination to favor the stranger over the friend on a scale from 0 (friend) to 100 (stranger) in response to human or an algorithm's advice in favor of either the friend or the stranger, where the human advisor is an impartial observer and the algorithm is modeled on impartial observers. The figure depicts means and standard errors, broken down by scenario, type of advisor, and nature of advice. \label{fig:Study1}}
		}{
			\includegraphics[width=.9\textwidth]{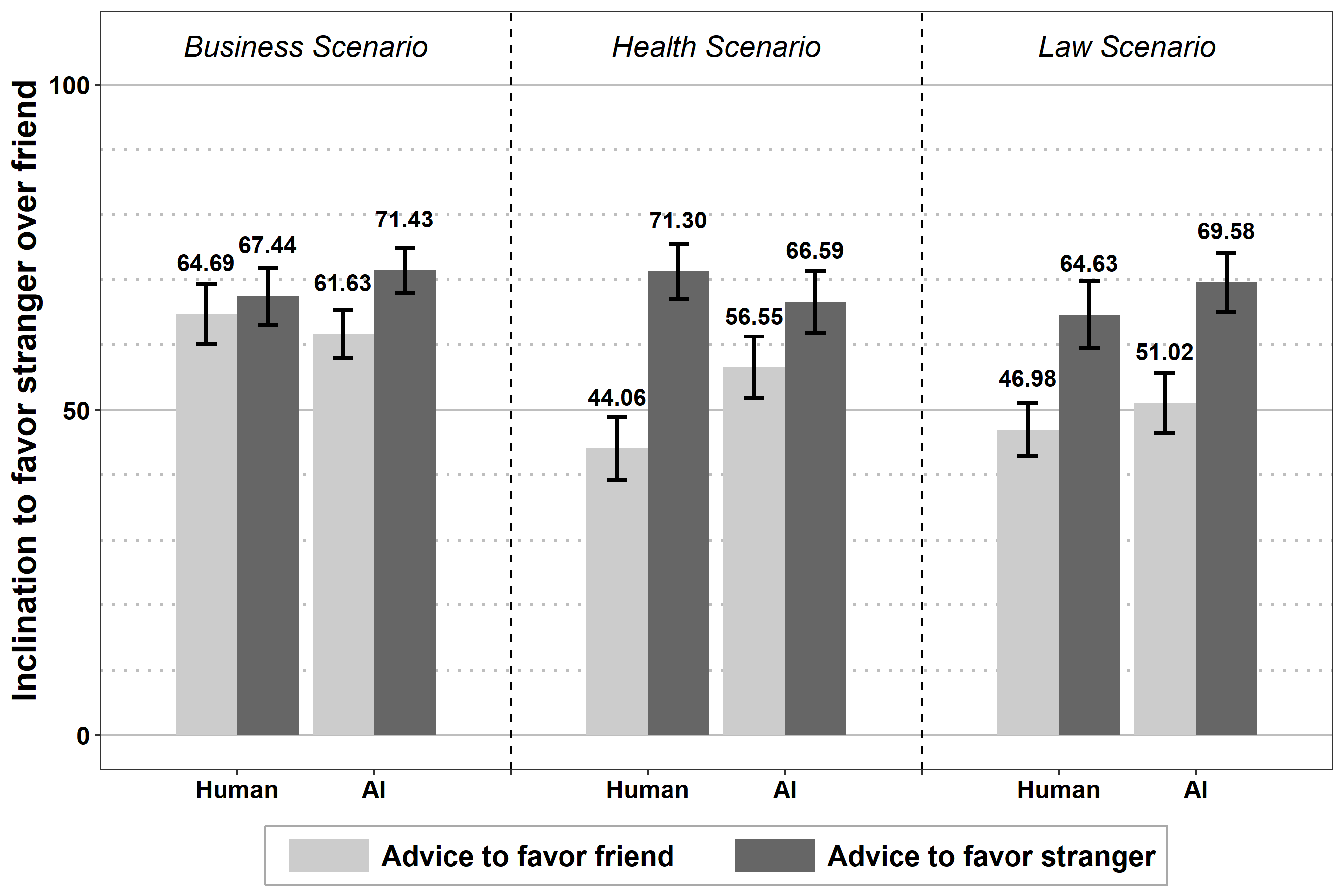}
		}
	\end{figure}

	Figure~\ref{fig:Study1} breaks the subjects' answers down by scenario, type of advisor, and nature of advice. Qualitatively, the subjects would rather decide in favor of the friend if advised that this was acceptable than if they were advised this was unacceptable, whether the advisor was human or an algorithm. The size of the difference varies somewhat between the scenarios and the types of advisor. The influence of the advice is apparently smaller in the business than in the other scenarios. We are interested, first, in whether ethical advice---both by the human advisor and the algorithm---leads the subjects to decide in favor of the stranger; second, in whether this effect differs depending on the type of advisor. To test the differences for significance, we ran regressions of the subjects' inclination to favor the stranger over the friend on the nature of advice and the type of advisor with scenario-specific random effects, as stated on AsPredicted.
	
	\begin{table}[tb]
		\ttabbox[.85\linewidth]{
			\caption{Influence of advice by AI-powered moral algorithm on moral judgment\label{tab:Study1}}
		}{
			\begin{tabularx}{\linewidth}{X*{3}{D{.}{.}{5.5}}}
				\toprule
				& \multicolumn{3}{c}{Advisor} \\
				\cmidrule(lr){2-4}
				& \multicolumn{1}{c}{AI}
				& \multicolumn{1}{c}{Human}
				& \multicolumn{1}{c}{Both} \\
				& \multicolumn{1}{c}{(1)}
				& \multicolumn{1}{c}{(2)}
				& \multicolumn{1}{c}{(3)} \\
				\midrule
				Advice to favor friend
				& -12.75^{***}
				& -16.03^{***}
				& -16.05^{***} \\
				& (3.52)
				& (3.75)
				& (3.67) \\
				AI-advisor
				&
				&
				& 1.32 \\
				&
				&
				& (3.63) \\
				AI-advisor $\times$ Advice to favor friend
				&
				&
				& 3.39 \\
				&
				&
				& (5.13) \\
				Intercept
				& 69.21^{***}
				& 67.74^{***}
				& 67.74^{***} \\
				& (2.58)
				& (3.65)
				& (3.38) \\
				\midrule
				Observations
				& 323
				& 310
				& 633 \\
				Groups (Scenario)
				& 3
				& 3
				& 3 \\
				Var: Intercept (Scenario)
				& 1.89
				& 18.41
				& 13.66 \\
				Var: Residual
				& 997.62
				& 1,\!088.76
				& 1,\!039.71 \\
				\bottomrule
			\end{tabularx}
			\footnotetext{
				\hspace{-2.15em} \textit{Note.} Regression of the inclination to favor the stranger over the friend on the nature of advice (decide for friend or stranger) and the type of advisor (AI-powered algorithm versus human), with scenario (business, health, and law) as random intercept. The inclination is measured on a scale ranging from 0 (for the friend) to 100 (for the stranger). Advice and AI-advisor are dummy variables.
				
				$^{***}\ p < 0.01$.
			}
		}
	\end{table}
	
	Columns~1 and 2 of Table~\ref{tab:Study1} list the results of separate regressions for either type of advisor. They show that advice to favor the friend, both when the advisor is human and when it is an algorithm, reduces decision-makers' inclination to decide in favor of the stranger, or makes them tend to decide in favor of the friend, across the three scenarios. Column~3 combines the data for both types of advisor to compare their effects. Again, the inclination to favor the stranger falls if the human advisor finds it acceptable to favor the friend. The coefficient on the interaction term between the algorithm as advisor and the advice to favor the friend is insignificant. Hence, there is no incremental effect of this type of advisor, and the effect of advice does not differ statistically between the human advisor and the algorithm. Likewise, the inclination does not differ depending on the type of advisor if the advice is against the friend.
	
	Having considered the effect of advice on decision-makers' inclination to decide in favor of the friend, another interesting question is how it would affect their actual decision. With an answer scale ranging from 0 (favor friend) to 100 (favor stranger), we took a score of less than 50 to indicate that the advisee would actually decide in favor of the friend, and thus derived a binary variable that captures the decision rather than the inclination. We see the percentage of advisees who decide to favor the friend rise from 23 to 43 if the human advisor finds this acceptable ($\chi^2 = 13.48$, $p < 0.01$). Likewise, it rises from 20 to 34 if the algorithm advises so ($\chi^2 = 6.57$, $p = 0.01$). The percentage does not differ between the human advisor and the algorithm, whether the advice was in favor of ($\chi^2 = 2.35$, $p = 0.13$) or against the friend ($\chi^2 = 0.09$, $p = 0.76$). Hence, the results for the inclination to decide turn out to also hold for the decisions.
	
	In summary, the algorithm's advice has an impact on decision-makers' inclination to favor the friend or stranger, and most likely on their actual decisions. Our decision-makers do not slavishly follow the algorithm's or the human advice, but the change in their inclination to favor the stranger or the friend depending on the nature of advice shows that they are influenced by it. Our data indicate neither algorithm appreciation nor algorithm aversion in the sense of the term used by \citet{DietvorstEtAl2015, DietvorstEtAl2018} and \citet{LoggEtAl2019}, both for the inclination and the presumptively resulting decision. Put differently, it appears that decision-makers do care about moral advice, but they do not care whether this advice comes from an impartial human advisor or an algorithm that imitates impartial human advisors, and our results do not reject the hypothesis that the algorithm's and the human advice have the same effect on the advisee.
	
	\section*{Study~2}
	
	The outcomes of our first study show that decision-makers heed moral advice from an artificial and a human advisor indifferently. They do not suggest algorithm aversion or distrust against algorithms compared to humans. In light of recent mixed evidence, this finding might not seem overly surprising. That said, despite this mixed evidence, the ``idea [of distrust against algorithms] is so prevalent that it has been adopted by popular culture and the business press'' \citep[][p.~90]{LoggEtAl2019}. This idea also informs ethics guidelines issued by governmental and non-governmental agencies and bodies, business associations, and companies \citep[for an overview, see][]{JobinEtAl2019}. For example, both the U.S. government and the European Union advocate and advance ``trustworthy AI'' because they are afraid that distrust might hinder the adoption and acceptance of AI \citetext{\citealp[e.g.,][p.~4]{EC2019}, \citealp[][Sec.~1]{EO13960}}.
	
	The link between trustworthiness and trust is not straightforward. We may trust or distrust others who then turn out worthy or unworthy of our trust, leaving us with situations where we properly place our trust or distrust in someone else or where we misplace our trust or distrust in them \citep{LevineEtAl2018}. Ability, benevolence, and integrity were identified as factors that lead us to expect that someone will turn out worthy of our trust and thus increase the likelihood that we trust them \citep{MayerEtAl1995}. In prior research, trust in AI was built on the ability of AI, namely to make correct factual predictions \citep{DietvorstEtAl2015, DietvorstEtAl2018, LoggEtAl2019}. Calls for transparent, explainable, or accountable AI, in turn, refer rather to \citeauthor{MayerEtAl1995}'s \citeyearpar{MayerEtAl1995} trustworthiness factor of integrity \citep{JobinEtAl2019}. For example, \citet{IBM2019} argues that ``we don't blindly trust those who can't explain their reasoning'' \citetext{p.~26}.
	
	Transparency is the most prevalent ethical principle in AI guidelines \citep{JobinEtAl2019}. It is considered a logical antecedent of trustworthiness. For instance, the \citet{IEEE2017} argues that ``transparency \textellipsis\ will allow a community to understand, predict, and appropriately trust the A/IS [autonomous and intelligent systems],'' and that ``transparency allows for trust to be maintained'' \citetext{p.~44}. Unlike explainability, which lawmakers propose but are reluctant to specify, transparency has been required by regulations for automated systems, like those scoring creditworthiness, since the 1970s \citep{WachterEtAl2017}. While transparency is seen as a means to make algorithms more trustworthy, there is considerable variation in what ethics guidelines claim should be made transparent, including how AI is used in decision-making, the source code, the training data, the legal regulations, the limitations, or the potential impact \citep{JobinEtAl2019}.
	
	Despite those claims, algorithms often remain untransparent because their proprietors refuse to disclose their functionality, or they are too complex to understand but by few specialists. Learning AI systems, in particular, are black boxes by design. Unlike rule-based AI systems, they develop their own rules to process information, which cannot meaningfully be interpreted by a human \citep{MittelstadtEtAl2016}. However, algorithms should then be made as transparent as possible and audited \citep{Mittelstadt2016}. For example, learning algorithms depend largely on their training data. Biases in those data will naturally affect the algorithm's decisions \citep{Mittelstadt2016}. The disclosure of (information about) the training data enables responsible users to assess the advice properly and decide what to make of it. It is therefore considered instrumental in making learning AI systems transparent and trustworthy \citep{EC2019}.
	
	If transparency makes an algorithm appear trustworthy and thus helps it earn its user's trust, users should be suspicious of algorithms they know nothing about. In our first study, the algorithm was described as mimicking human decisions, based on the judgments of impartial human advisors. This description suggests that the algorithm's advice resembles the advice of those human advisors. While it is hard to tell how human advisors make their judgments, this algorithm stands comparison with them. In this sense, the disclosure of information about the training data made the algorithm of our first study transparent. Conversely, the lack of any such information makes an algorithm less transparent, compared to this benchmark. If users distrust untransparent algorithms, they should be more reluctant to heed advice from such an algorithm relative to the algorithm from our first study and hence make their judgments more independently.
	
	\subsection*{Method}
	
	We ran another experiment on CloudResearch to investigate the impact of information about how the algorithm formed its advice. The vignettes employed the same wording as those featuring the algorithm as an advisor in our first study. They differed only in stating that it was unknown what the algorithm's advice was based on rather than that it was based on the judgments of impartial human advisors. As opposed to our first study, there is no meaningful parallel condition with an untransparent human advisor. Humans are notoriously untransparent, and therefore the decision-makers did not learn anything about the human advisor in our first study that could be concealed. Instead, the untransparent algorithm is a viable alternative substitute for the human advisor, and the advisees' response to the human advice in our first study remains a valid benchmark to assess their response to the untransparent algorithm in the second.
	
	\subsection*{Results}
	
	A total of 396 subjects took our survey. 309 of these passed the comprehension test. 35 percent of the 309 subjects were male. The subjects' age ranged from 17 to 91 years, with a mean of 43.9 years. The $2 \times 3$ factorial design, where the algorithm advised either in favor of or against the friend in one of three scenarios, created six experimental conditions. The subjects were randomly assigned to these conditions. Like in our first study, we focus on how much they agreed with the statement that they would rather decide in favor of the stranger or the friend. If the advisees care about transparency, they are reluctant to follow advice from an opaque algorithm, and their agreement should not differ depending on the nature of advice. If it differs nonetheless, we are led to conclude that opaqueness does not bother them, and that the effect of the algorithm's advice that we observed in our first study is therefore not driven by its relative transparency.
	
\begin{figure}[t]
		\ffigbox[\FBwidth]{
			\caption{Inclination to favor the stranger over the friend on a scale from 0 (friend) to 100 (stranger) in response to an algorithm's advice in favor of either the friend or the stranger, where nothing is known about how the algorithm works. The figure depicts means and standard errors, broken down by scenario and nature of advice. \label{fig:Study2}}
		}{
			\includegraphics[width=.9\textwidth]{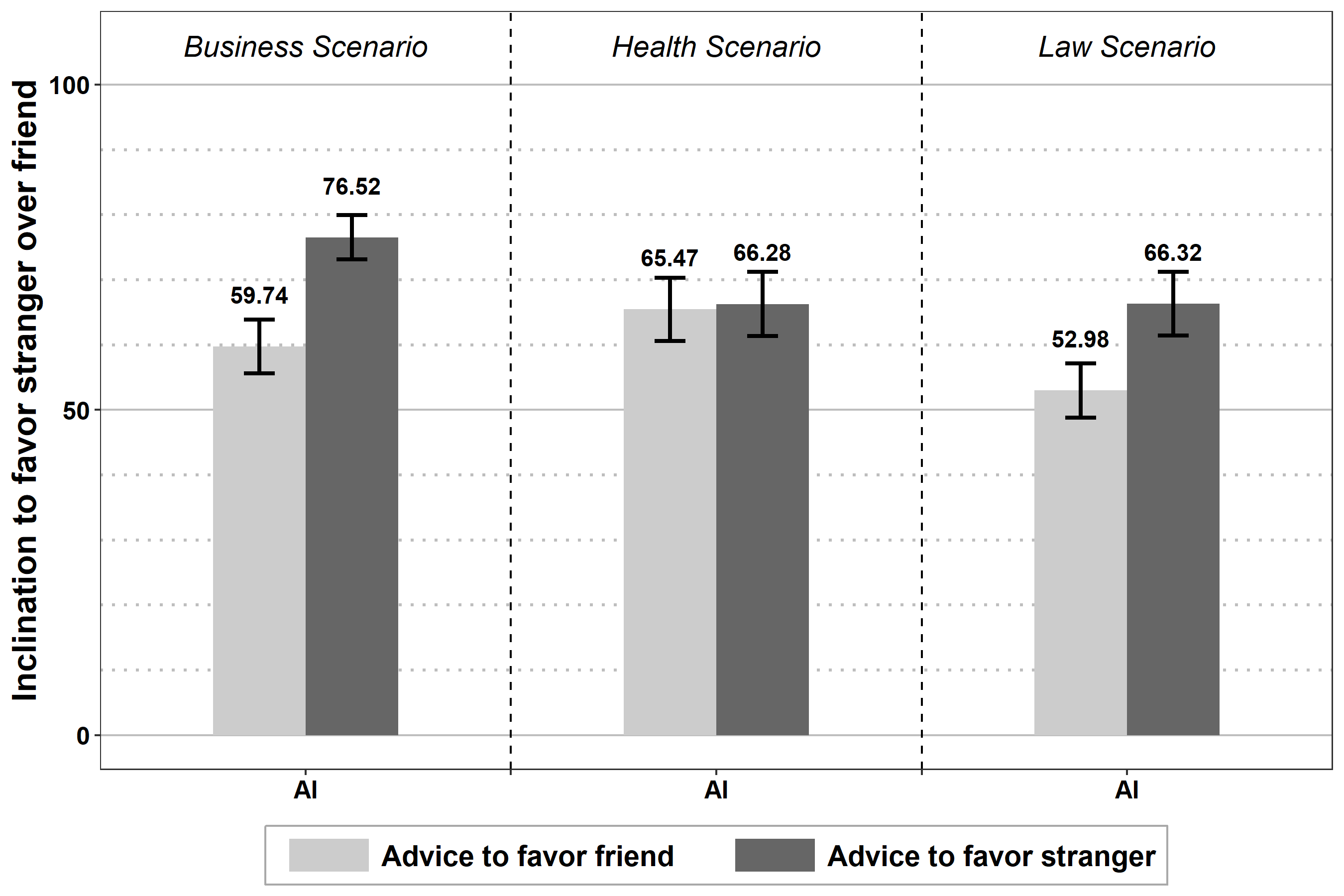}
		}
	\end{figure}

	Figure~\ref{fig:Study2} depicts the subjects' answers, broken down by nature of advice and scenario. The results turn out to be qualitatively similar to those of our first study. As recruiters in the business scenario, the subjects leaned more toward hiring the stranger when advised so by the algorithm; as prosecutors in the law scenario, they tended more to raid the suspicious financial services firm run by the friend than the one run by the stranger in this case. Only for the health officials compiling the waiting list for donated kidneys in the health scenario, the algorithm's advice makes little difference. Their agreement with deciding in favor of the stranger is similar regardless of the nature of advice, which suggests that they are less susceptible to it. Taking together the outcomes of the three scenarios, however, advisees appear to heed the advice although the algorithm is opaque, and they are therefore uncertain about what its advice is based on.
	
	\begin{table}[tb]
		\ttabbox[.9\linewidth]{
			\caption{Influence of advice by AI-powered opaque algorithm on moral judgment\label{tab:Study2}}
		}{
			\begin{tabularx}{\linewidth}{X*{3}{D{.}{.}{5.5}}}
				\toprule
				& \multicolumn{3}{c}{Advisor} \\
				\cmidrule(lr){2-4}
				& \multicolumn{1}{c}{Opaque}
				&
				& \\
				& \multicolumn{1}{c}{AI}
				& \multicolumn{1}{c}{Human}
				& \multicolumn{1}{c}{Both} \\
				& \multicolumn{1}{c}{(1)}
				& \multicolumn{1}{c}{(2)}
				& \multicolumn{1}{c}{(3)} \\
				\midrule
				Advice to favor friend
				& -11.01^{***}
				& -16.03^{***}
				& -15.97^{***} \\
				& (3.56)
				& (3.75)
				& (3.65) \\
				AI-advisor
				&
				&
				& 2.27 \\
				&
				&
				& (3.68) \\
				AI-Advisor $\times$ Advice to favor friend
				&
				&
				& 5.07 \\
				&
				&
				& (5.17) \\
				Intercept
				& 70.17^{***}
				& 67.74^{***}
				& 67.73^{***} \\
				& (3.19)
				& (3.65)
				& (3.55) \\
				\midrule
				Observations
				& 309
				& 310
				& 619 \\
				Groups (Scenario)
				& 3
				& 3
				& 3 \\
				Var: Intercept (Scenario)
				& 11.36
				& 18.41
				& 17.32 \\
				Var: Residual
				& 979.07
				& 1,\!088.76
				& 1,\!032.14 \\
				\bottomrule
			\end{tabularx}
			\footnotetext{
				\hspace{-2.15em} \textit{Note.} Regression of the inclination to favor the stranger over the friend on the nature of advice (decide for friend or stranger) and the type of advisor (AI-powered algorithm versus human), with scenario (business, health, and law) as random intercept. The inclination is measured on a scale ranging from 0 (for the friend) to 100 (for the stranger). Advice and AI-advisor are dummy variables.
				
				Column~1 is based on Study~2. Column~2 is the same as Column~2 of Table~\ref{tab:Study1}. Column~3 is based on the data of both Study~2 (opaque algorithm) and Study~1 (human advisor).
				
				$^{***}\ p < 0.01$.
			}
		}
	\end{table}
	
	To test for potential differences, we first ran a regression of the subjects' inclination to favor the stranger over the friend on the nature of advice, again with scenario-specific random effects. The results in Column~1 of Table~\ref{tab:Study2} show that the algorithm's advice to favor the friend reduces this inclination. This suggests a significant influence of the algorithm although it is opaque. To complete the picture, the effect of the algorithm's advice needs to be benchmarked against that of human advice---i.e., the response of decision-makers who were told that they were advised by an impartial human advisor in our first study. Column~2 reproduces Column~2 of Table~\ref{tab:Study1} for convenience. Combining the data, neither the coefficient on the interaction term nor on the algorithm as advisor in Column~3 differ significantly from zero. Hence, like in our first study, the effect of the algorithm's advice does not statistically differ from that of the human advice.
	
	To complement our analysis, we derived, in the same way as in our first study, a binary variable that captures the advisees' actual decisions rather than their inclination to decide in favor of or against the friend. Looking at this variable, the percentage of advisees who would decide in favor of the friend is 21 if the algorithm advises against the friend, as opposed to 32, if it advises in favor of the friend. This difference is marginally significant ($\chi^2 = 3.60$, $p = 0.058$), and like in our first study, this result can be taken to indicate that the algorithm's influence on decision-makers' inclination to decide in favor of the stranger or the friend carries over to their actual decisions. For comparison, we recall that the percentages are similar in our first study, where 23 or 43 percent of the advisees would decide in favor of the friend, depending on what the human advice says, and 20 and 34 percent, depending on the transparent algorithm's advice.
	
	This outcome is not intuitive and it challenges the common belief that transparency creates trust. A potential explanation is motivated reasoning. People tend to collect and evaluate information in the light of the conclusions that they want to reach, and they disregard, overlook, or reinterpret conflicting information \citep{Kunda1990, Gilovich1991}. In our experiment, the subjects faced a moral dilemma that demanded a burdensome decision between friendship and duty. It was therefore convenient for them to follow the algorithm's advice to relieve this burden. The vignette made the opaqueness of the algorithm salient, but the advisees still knew that the algorithm was there to ``tell whether it [was] ethically acceptable to decide in favor of a friend in such situations.'' It is conceivable that they filled the void and just assumed the algorithm was trustworthy although nothing was said to that effect. This conjecture motivates our next study.
	
	Summing up, decision-makers trust moral advice from more and less transparent algorithms alike. While they do not each decide as advised, of course, their decisions are clearly influenced by the advice. We cannot rule out that the response to advice coming from an opaque algorithm is the same as to advice coming from an impartial human advisor. Again, we observe neither algorithm appreciation nor aversion relative to human advice. It is important to note that this finding does not invalidate the ethical argument for transparency to increase trustworthiness. It does suggest, though, that decision-makers attach less value to transparency or trustworthiness than commonly thought. Empirically, they blindly follow advice from an algorithm they know little about. Practically, this observation casts doubt on whether a decision-maker, as the human in the loop, can be expected to effectively control the algorithm's decision in augmented decision-making.
	
	\section*{Study~3}
	
	Starting from inconclusive evidence about trust in advice from AI-powered algorithms, we found in our first study that decision-makers trust ethical advice from an algorithm and an impartial human advisor alike. The algorithm was described as imitating decisions of impartial human advisors, though, making its training data quite transparent. Conceptually, transparency is an antecedent of trustworthiness, and a trustworthy algorithm is more likely to be trusted. This is arguably why transparency is the most prevalent principle in AI guidelines \citep{JobinEtAl2019}. Unfortunately, transparency is often hard to attain \citep{Mittelstadt2016}. To test how much transparency matters, we made the algorithm opaquer and told the subjects in our second study that it was unknown what the algorithm's advice was based on. Interestingly, they turned out to follow the algorithm's advice both when it was more and when it was less transparent.
	
	Having withheld information which suggests that the algorithm is trustworthy (i.e., that it imitates impartial human advisors), the next step to challenge users' trust is to provide information which casts doubt on the algorithm's trustworthiness. In our first study, trust in the algorithm's moral judgment was derived from its training data. Obviously, there is little reason to distrust the moral judgment of impartial human advisors. It then seems intuitive to infer that the algorithm's resemble those advisors' decisions, making the algortihm's advice trustworthy by the same token, and our subjects followed indeed the algorithm's advice. Suppose now that the algorithm is instead designed to imitate decisions which are based on the moral judgments of people who have presumably acted immorally, such as convicted criminals. Will advisees still heed the algorithm's advice, or will they now disregard it and decide independently of the advice?
	
	It is not straightforward to conclude that the moral judgment of criminals is untrustworthy. First, we imply that the advisee shares the norms of the (unspecified) legal system under which the criminals have been convicted and that this legal system has rightly convicted them. That is, the convicts advising her are indeed criminals and the deeds that led to their convictions were not only illegal but also immoral. We assume that advisees take a conviction to indicate a moral transgression. Second, a moral transgression does not necessarily result from a lack of moral judgment. A criminal might know well what is morally right to do in some situation but do the wrong thing nonetheless. She would then be a poor role model, but her moral judgment would be intact, and she would therefore be perfectly qualified to give moral advice. (``Do as I say, not as I do.'') Is it reasonable to assume that a criminal's moral judgment is trustworthy, on average?
	
	Psychological research shows that individuals who load high on the dark triad personality traits \citep{PaulhusWilliams2002} show an impaired moral judgment and a lower level of moral development than the average population \citep{CampbellEtAl2009, JonasonEtAl2015, Blair2017}. These traits are particularly pronounced among criminal offenders, who are often the subjects of such research. While juvenile offenders' moral judgment is clearly impaired \citep{StamsEtAl2006}, the evidence for adult criminals is more mixed. However, recent studies did find systematic differences \citep{KoenigsEtAl2012, YoungEtAl2012, LahatEtAl2015}. For example, psychopaths attach lower relevance to fairness, authority, and others' suffering \citep{JonasonEtAl2015}, and they have lower reservations against inflicting harm on others \citep{KoenigsEtAl2012}. Overall, this research suggests that convicts' moral judgment differs from the general population.
	
	This evidence gives reason to assume that many perceive the moral judgment of convicts as biased. Convicts are thus considered a negative selection of moral advisors, if not relative to the average population, then certainly to the impartial observers from our first study. Practically, the discrimination against ex-offenders in education, employment, and housing shows deep-seated distrust \citep{SokoloffSchenckFontaine2017, EvansEtAl2019, SugieEtAl2020}. Of course, this distrust is arguably driven by fear of continued immoral behavior, which results from poor moral judgment or practical judgment or both. That said, people do not necessarily distinguish between moral and practical judgment. Instead, they will arguably find convicts' moral judgment untrustworthy just because they find convicts untrustworthy. Hence, one would expect that decision-makers disregard advice from convicts, and that it has little effect on their decisions.
	
	\subsection*{Method}
	
	Along the lines of our first study, we ran another experiment on CloudResearch, where the algorithm's advice was described as based on the ethical judgments of convicted criminals rather than impartial human advisors. Other than that, we used the exact same vignettes as in the first two studies, where the decision-maker (e.g., the recruiter) was confidentially advised by an algorithm on whether to favor the friend or the stranger. To benchmark the impact of advice by the criminals-trained algorithm against that of a human advisor, we ran another condition where a human convicted criminal replaced the impartial human advisor from our first study. This condition also allows us to validate our argument that decision-makers distrust the moral judgment of convicted criminals. If we are right in assuming that a criminal record undermines an advisor's trustworthiness in the eyes of the advisee, we shall see the advisees disregard the convict's advice.
	
	\subsection*{Results}
	
	We collected answers from a total of 796 subjects for this study. 651 answered the comprehension question correctly. 52 percent of these subjects were male, and their average age was 45.9 years, within a range from 17 to 90 years. Like our first study, the experiment has a $2 \times 2 \times 3$ factorial design, where either a convicted criminal or an algorithm who imitates the decisions of convicted criminals gives the advice that it is either acceptable or unacceptable for the decision-maker to favor the friend in one of our three scenarios. We ran first the conditions with the algorithm as advisor and subsequently those with the human advisor. Within either run, the subjects were randomly assigned to the six conditions. The focal variable is again how much the subjects tended to decide in favor of the stranger as opposed to the friend in their role as a recruiter, prosecutor, or health official who compiles the waiting list for donated kidneys.
	
\begin{figure}[tb]
		\ffigbox[\FBwidth]{
			\caption{Inclination to favor the stranger over the friend on a scale from 0 (friend) to 100 (stranger) in response to human or an algorithm's advice in favor of either the friend or the stranger, where the human advisor is a convicted criminal and the algorithm is modeled on convicted criminals. The figure depicts means and standard errors, broken down by scenario, type of advisor, and nature of advice. \label{fig:Study3}}
		}{
			\includegraphics[width=0.9\textwidth]{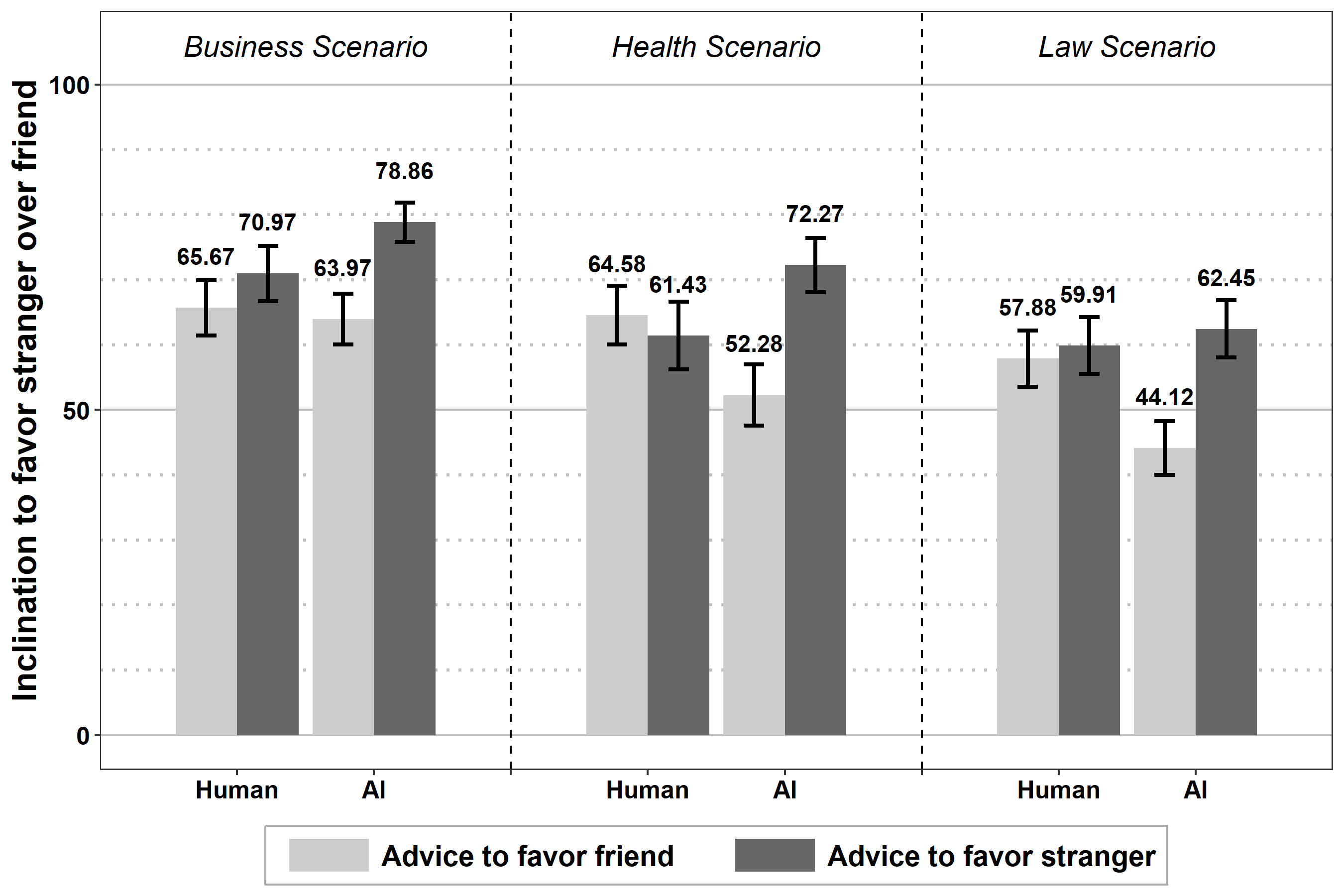}
		}
	\end{figure}

	Figure~\ref{fig:Study3} depicts the results, which differ qualitatively from those in our first and second studies in Figures~\ref{fig:Study1} and \ref{fig:Study2}. On the one hand, the decision-makers who were advised by the human advisor (i.e., a convicted criminal) showed a similar inclination to decide in favor of the friend, regardless of whether the advice said that it was acceptable or unacceptable to favor the friend. Put differently, the descriptive statistics suggest that they considered ethical advice coming from a criminal to be untrustworthy and consequently disregarded it, as one would expect. On the other hand, the decision-makers who were advised by the algorithm instead still tended to favor the friend more if the advice said that this was acceptable and less so if it said that this was unacceptable. Hence, it turns out that they heeded the algorithm's advice although that algorithm was introduced as being trained on the moral judgments of convicted criminals.
	
	\begin{table}[tb]
		\ttabbox[\linewidth]{
			\caption{Influence of advice by AI-powered criminal algorithm on moral judgment\label{tab:Study3}}
		}{
			\begin{tabularx}{\linewidth}{X*{5}{D{.}{.}{5.5}}}
				\toprule
				& \multicolumn{5}{c}{Advisor} \\
				\cmidrule(lr){2-6}
				&
				&
				& \multicolumn{1}{c}{Criminal}
				&
				& \multicolumn{1}{c}{Criminal} \\
				& \multicolumn{1}{c}{Criminal}
				&
				& \multicolumn{1}{c}{AI \&}
				&
				& \multicolumn{1}{c}{AI \&} \\
				& \multicolumn{1}{c}{AI}
				& \multicolumn{1}{c}{Criminal}
				& \multicolumn{1}{c}{Criminal}
				& \multicolumn{1}{c}{Human}
				& \multicolumn{1}{c}{Human} \\
				& \multicolumn{1}{c}{(1)}
				& \multicolumn{1}{c}{(2)}
				& \multicolumn{1}{c}{(3)}
				& \multicolumn{1}{c}{(4)}
				& \multicolumn{1}{c}{(5)} \\
				\midrule
				Advice to favor friend
				& -17.52^{***}
				& -1.59
				& -1.55
				& -16.03^{***}
				& -15.73^{***}\\
				& (3.29)
				& (3.64)
				& (3.45)
				& (3.57)
				& (3.55) \\
				AI-advisor
				&
				&
				& 7.03^{**}
				&
				& 3.52 \\
				&
				&
				& (3.47)
				&
				& (3.54) \\
				AI-Advisor $\times$ Advice
				&
				&
				& -15.89^{***}
				&
				& -1.73 \\
				\hspace{1em} to favor friend
				&
				&
				& (4.92)
				&
				& (4.98) \\
				Intercept
				& 71.16^{***}
				& 64.38^{***}
				& 64.19^{***}
				& 67.74^{***}
				& 67.68^{***} \\
				& (5.51)
				& (3.32)
				& (4.53)
				& (3.65)
				& (4.71) \\
				\midrule
				Observations
				& 321
				& 330
				& 651
				& 310
				& 631 \\
				Groups (Scenario)
				& 3
				& 3
				& 3
				& 3
				& 3 \\
				\textls[-60]{Var: Intercept (Scenario)}
				& 74.96
				& 13.04
				& 43.43
				& 18.41
				& 47.13 \\
				Var: Residual
				& 869.34
				& 1,\!090.84
				& 981.88
				& 1,\!088.76
				& 977.23 \\
				\bottomrule
			\end{tabularx}
			\footnotetext{
				\hspace{-2.15em} \textit{Note.} Regression of the inclination to favor the stranger over the friend on the nature of advice (decide for friend or stranger) and the type of advisor (AI-powered algorithm versus human), with scenario (business, health, and law) as random intercept. The inclination is measured on a scale ranging from 0 (for the friend) to 100 (for the stranger). Advice and AI-advisor are dummy variables.
				
				Columns~1--3 are based on Study~3; Column~4, on Study~1 (Column~2 of Table~\ref{tab:Study1}). Column~5 is based on the data of both Study~3 (criminal algorithm) and Study~1 (human advisor).
				
				$^{**}\ p < 0.05$. $^{***}\ p < 0.01$.
			}
		}
	\end{table}
	
	Like in the previous two studies, we ran regressions with scenario-specific random effects of the decision-makers' inclination to favor the stranger over the friend on the type of advisor and the nature of advice to test these qualitative differences for significance. Table~\ref{tab:Study3} shows the results of separate regressions for the algorithm mimicking convicted criminals (Column~1) and the convicted criminal as human advisor (Column~2) as well as a regression on the whole data set (Column~3). Like the convicted criminal, the impartial human advisor from our first study is, again, a meaningful benchmark for the influence of the convicts-trained algorithm's advice. Column~4 reproduces the results listed in Column~2 of Table~\ref{tab:Study1} for convenience. Column~5 lists the results of the regression on the data for the decision-makers receiving advice from the human advisor from our first study and from the algorithm from our third study combined.
	
	The results in Columns~1 and 2 show that advice to favor the friend reduces the inclination to favor the stranger if it comes from the convicts-trained algorithm but not the convict. Accordingly, Column~3 reports a negative coefficient on the interaction term but not the advice to favor the friend. Incidentally, advice to favor the stranger by the algorithm increases this inclination relative to the same advice coming from the convict, as can be seen from the significant coefficient on the algorithm as advisor. These results argue for an impact of advice by the algorithm but not the convict. When we compare the algorithm to the impartial human advisor from our first study, we observe the same pattern in Column~5 as in the previous two studies (Column~3 of Table~\ref{tab:Study1} and \ref{tab:Study2}, respectively): The influence of the algorithm does not differ from that of the impartial human advisor, although the algorithm is trained on convicts.
	
	As in the previous two studies, the results for the inclination to decide for the friend or stranger also hold for the actual decision. Translating the inclination score into a decision as before, the percentage of decision-makers who favor the friend is almost the same, whether the convict finds it acceptable or unacceptable to favor the friend (27 and 28). By contrast, if the convicts-trained algorithm advises in favor of the friend rather than the stranger, the percentage is 36 compared to 15, and thus significantly higher ($\chi^2 = 17.72$, $p < 0.01$). Recalling the percentages of decision-makers leaning toward these decisions with advice from the impartial human advisor from our first study, which are 43 and 23, the difference in percentage points is about the same. Hence, the influence of the convicts-trained algorithm resembles that of the impartial human advisor, which matches the insignificant effect of the interaction term in Column~5.
	
	In summary, decision-makers are influenced by an algorithm's advice even if suspicious training data give them reason to distrust it. Indeed, the influence of that algorithm does not statistically differ from that of an impartial human advisor. In turn, the influence of a criminal human advisor is zero, ruling out the possibility that decision-makers are always influenced by advice and never care who the advisor is. Hence, we undermined the algorithm's trustworthiness by both concealing information that could argue for its trustworthiness and revealing information that argues against it. Neither does the lack of transparency reduce trust, nor does transparency immunize advisees against advice from an untrustworthy algorithm. It seems that, in addition to transparency of information about the training data, digital literacy is needed for users to benefit from this transparency and make use of their information about the algorithm.
	
	\section*{Further Results}
	
	Having found that decision-makers are influenced by ethical advice, both when it comes from a human advisor and when it comes from an algorithm, we wonder how they perceive the role of advice in their judgements and decisions. To address this question, we posed a post-experimental question to ask our subjects, first, whether they would make the same judgment if there were no moral advice. Second, we asked them the same question about ``most other participants.'' Table~\ref{tab:Influence} summarizes the answers, broken down by the type of advisor and the nature of advice. We note that about three quarters of the subjects, with little variation among the experimental conditions, claimed that they would make the same judgment, which suggests that they felt hardly influenced by the advice. Conversely, more than half of them believed that the other participants were influenced and therefore would make a different judgment without advice.
	
	\begin{table}
		\ttabbox[.8\linewidth]{
			\caption{Participants' assessment of influence of advice \label{tab:Influence}}
		}{
			\begin{tabularx}{\linewidth}{L*{4}{>{\hsize=.33\hsize}C}}
				\toprule
				& \multicolumn{2}{c}{AI advisor}
				& \multicolumn{2}{c}{Human advisor} \\
				\cmidrule(lr){2-3} \cmidrule(lr){4-5}
				& \multicolumn{2}{c}{Advice to favor}
				& \multicolumn{2}{c}{Advice to favor} \\
				\cmidrule(lr){2-3} \cmidrule(lr){4-5}
				& Friend
				& Stranger
				& Friend
				& Stranger \\
				\midrule
				Influence on decision-maker
				& 26\%
				& 21\%
				& 25\%
				& 23\% \\
				Influence on others
				& 51\%
				& 60\%
				& 57\%
				& 56\% \\
				\bottomrule
			\end{tabularx}
			\footnotetext{
				\hspace{-2.15em} \textit{Note.} The participants were asked to confirm or refute the following statements: ``If there were no ethical advice, I would make a different judgment''; ``If there were no ethical advice, most other participants would make a different judgment.'' The numbers in the various AI-treatments were virtually identical and we therefore pooled them together. 
			}
		}
	\end{table}
	
	These observations are striking in two regards. On the one hand, it is noteworthy how few subjects considered themselves susceptible to advice despite the large and significant effect that our three studies establish. On the other other hand, they consider others much more susceptible to it than themselves. The same kind of self-defeating reasoning has been found for morality. Departing from the fact that people consider themselves more selfless, kind, and generous than others, \citet{EpleyDunning2000} show that this is not because they underestimate others, but because they overestimate themselves, while their assessment of others is quite accurate. The subjects' answers point to a considerable potential for self-deception about the influence of advice, and particularly advice by algorithms. There is a risk that decision-makers continue to believe that they own their decisions, although they largely adopt them from machines.
	
	To examine the influence of advice on moral judgements and decisions, we intentionally raised the ethical question of whether to favor a friend over a stranger. While this question allows for a role of moral advice, the answer might also hinge on the decision-maker's personal morality. To consider the potential effect of personal morality, we included a post-experimental question about ethical self-assessment, which asked the subjects how moral they considered themselves relative to other subjects. They answered on a scale ranging from 0 to 100, where a score of above 50 identified the respondent as more ethical than others; below 50, as less ethical. The answers actually range from 0 to 100, with a mean of 68.41 and a standard deviation of 19.61. We added moral self-assessment as a covariate to our regressions to test whether it is associated with the inclination to favor the stranger over the friend and to isolate the influence of ethical advice.
	
	The results, which are reported in Table~\ref{tab:Appendix_Reg} in the appendix, show a significant positive coefficient on self-assessed morality. Hence, decision-makers who considered themselves morally superior to others tend more to decide in favor of the stranger. Albeit not crucial for our findings, the decision in favor of the stranger thus turns out to be the morally superior choice in the eyes of our subjects. The focal effects of the algorithm, the advice, and their interaction do not differ qualitatively from those without ethical self-assessment in Tables~\ref{tab:Study1}--\ref{tab:Study3}, confirming our results. Incidentally, the aforementioned mean response of 68.41 reveals that our subjects assessed themselves to be more moral than the average. This is another example of the ``better-than-average effect,'' which is well-documented in the social psychology literature \citep{KoellingerEtAl2007, MerkleWeber2011}, and which motivated \citeauthor{EpleyDunning2000}'s \citeyearpar{EpleyDunning2000} above-cited study.
	
	Additionally, we asked our subjects another question about their ethical mindset as well as some questions about their attitude to artificial intelligence. Specifically, we used the standard trolley problem by \citet{Foot1967} to explore whether they were rather outcome-minded or rule-minded \citep{CornelissenEtAl2013}, and added their responses to our regressions. The results in Table~\ref{tab:Appendix_Reg} in the appendix reveal that outcome-minded decision-makers, who are identified by diverting the run-away trolley to kill one person and save five, tended slightly more to decide in favor of the friend than rule-minded subjects. Keeping in mind that favoring the stranger over the friend is considered more moral by the subjects, it is intuitive that outcome-minded people are more willing to trade off morality for friendship; however, the effect is not significant. Likewise, openness to artificial intelligence does not play a role for our findings.
	
	\section*{Conclusion}
	
	We ran three experiments on trust in algorithms in the moral domain, where the subjects took the role of a decision-maker who faces a dilemma between friendship and duty. We manipulated the trustworthiness of the algorithm by concealing information about its training data, thus making it less transparent, and by revealing information that suggested it was biased. We found, first, that the algorithm's advice influences users' decisions like advice from an impartial human advisor. If it encourages a decision in favor of the friend, users are more inclined to decide in favor of the friend; if it discourages that decision, they are less inclined so. Second, users care little about how trustworthy the algorithm is. Its influence is almost the same whether it is presented as imitating impartial human advisors, as a black box, or as mimicking convicted criminals. By contrast, decision-makers do disregard advice from a human convicted criminal.
	
	Our findings contribute to the growing literature on the ethical design of AI. It is commonplace that AI tends to be distrusted and that transparent and thus trustworthy AI is needed to reap the societal benefits AI can bring. This, however, is not what we find. Our data suggest that people follow AI-generated as much as much as human advice; that they do not bother about untransparency; that they follow AI-generated advice even in the presence of suspicous information; that they (over)confidently believe others are more susceptible to the algorithm's influence than themselves. These empirical observations do not invalidate the ethical argument for transparent and trustworthy AI. They do indicate, though, that more than that is needed for a responsible use of AI. While users can realize, by trail-and-error, that AI can err to become more diligent \citep{DietvorstEtAl2015}, we feel it is better to improve digital literacy \citep{BurtonEtAl2020}.
	
	Moreover, our research extends the ongoing debate in social sciences on whether and when people trust or distrust algorithms. \citet{DietvorstEtAl2015, DietvorstEtAl2018} presented evidence for algorithm aversion. \citet{LoggEtAl2019}, in contrast, found that advisees tend to appreciate AI-generated relative to human advice. Adding to this inconclusive prior evidence, we observe neither algorithm appreciation nor algorithm aversion about our moral dilemma. Instead, the results of our experiment suggest that the impact of AI-generated and human moral advice are very similar. (Technically, we see appreciation of the algorithm that mimics convicts relative to the human convict, but not the impartial human advisor.) As evidence keeps accumulating on both sides, along with evidence like ours, which argues for neither side, it seems that neither algorithm appreciation nor algorithm aversion generally prevails, but that multiple factors matter.
	
	This paper naturally has limitations that inivite further research. We considered a dilemma between friendship and duty to study trust in AI-powered algorithms. We varied the scenario to cover occupational, medical, and legal decisions, but it is easy to conceive further scenarios and dilemmas. While we believe that our findings generalize, it would be desirable to see them stand up to variation on both counts. We also confined ourselves to learning as opposed to rule-based AI. Learning machines have strenghts, including their flexiblity and scalability. However, their trustworthiness is limited by their training data and other factors. Rule-based machines, in turn, can be rendered fully transparent and users can thus check whether they agree with the rules. There have been attempts to create rule-based AI-advisors in the moral domain \citep{LaraDeckers2020}, and it would be interesting to see similar research on them.
	
	AI-powered advisors have a tremendous potential of improving decision-making. On the one hand, it is good news that ethical advice from algorithms is accepted. On the other hand, it is worrisome how little users reflect on such advice, even when they are cautioned against it. This is also a caveat against the human-in-the-loop approach: It makes us feel better about the resulting decision, but it cannot mitigate this risk if the human in the loop succumbs to the temptation of trusting the algorithm too readily. This overtrust creates a risk for decision-making to be corrupted by flawed algorithms. In a future with AI-powered assistants supporting us in all areas of life, we cannot count on government or other regulation alone to ensure that these are trustworthy, and we do not want to wait for users to find out that algorithms can err. Instead, we need to improve digital literacy and train them to use algorithms responsibly.
	
	\clearpage
	
	\singlespacing
	
	\bibliography{Refs}
	
	\clearpage
	
	\appendix
	
	\section*{Further Data}
	
	Table~\ref{tab:PEQ} summarizes the answers to the post-experimental questions and demographic characteristics by study and treatment.
	
	\begin{center}
		\begin{turn}{90}
			\begin{minipage}[][][c]{17.5cm}
				\ttabbox[17.5cm]{
					\caption{Demographic characteristics and responses to post-experimental questions \label{tab:PEQ}}
				}{
					\begin{tabularx}{17.5cm}{X*{10}{D{.}{.}{2.2}}} 
						\toprule
						& \multicolumn{4}{c}{Study~1}
						& \multicolumn{2}{c}{Study~2}
						& \multicolumn{4}{c}{Study~3} \\
						\cmidrule(lr){2-5} \cmidrule(lr){6-7} \cmidrule(lr){8-11}
						& \multicolumn{2}{c}{Human advisor}
						& \multicolumn{2}{c}{AI advisor}
						& \multicolumn{2}{c}{AI advisor}
						& \multicolumn{2}{c}{Human advisor}
						& \multicolumn{2}{c}{AI advisor} \\
						\cmidrule(lr){2-3}  \cmidrule(lr){4-5} \cmidrule(lr){6-7} \cmidrule(lr){8-9} \cmidrule(lr){10-11}
						& \multicolumn{2}{c}{Advice to favor}
						& \multicolumn{2}{c}{Advice to favor}
						& \multicolumn{2}{c}{Advice to favor}
						& \multicolumn{2}{c}{Advice to favor}
						& \multicolumn{2}{c}{Advice to favor} \\
						\cmidrule(lr){2-3}  \cmidrule(lr){4-5} \cmidrule(lr){6-7} \cmidrule(lr){8-9} \cmidrule(lr){10-11}
						& \multicolumn{1}{c}{Friend}
						& \multicolumn{1}{c}{\textls[-75]{Stranger}}
						& \multicolumn{1}{c}{Friend}
						& \multicolumn{1}{c}{\textls[-75]{Stranger}}
						& \multicolumn{1}{c}{Friend}
						& \multicolumn{1}{c}{\textls[-75]{Stranger}}
						& \multicolumn{1}{c}{Friend}
						& \multicolumn{1}{c}{\textls[-75]{Stranger}}
						& \multicolumn{1}{c}{Friend}
						& \multicolumn{1}{c}{\textls[-75]{Stranger}} \\
						\midrule
						Ethical self-
						& 65.57
						& 69.41
						& 66.31
						& 71.20
						& 67.95
						& 68.11
						& 67.72
						& 68.82
						& 68.53
						& 70.35 \\
						\hspace{1em} assessment
						& (21.02)
						& (19.47)
						& (18.65)
						& (19.08)
						& (19.80)
						& (18.92)
						& (20.20)
						& (21.44)
						& (19.24)
						& (17.80) \\
						Divert trolley
						& \multicolumn{1}{c}{85\%}
						& \multicolumn{1}{c}{88\%}
						& \multicolumn{1}{c}{89\%}
						& \multicolumn{1}{c}{84\%}
						& \multicolumn{1}{c}{92\%}
						& \multicolumn{1}{c}{88\%}
						& \multicolumn{1}{c}{92\%}
						& \multicolumn{1}{c}{91\%}
						& \multicolumn{1}{c}{89\%}
						& \multicolumn{1}{c}{91\%} \\
						Excited about
						& 3.00
						& 2.87
						& 2.74
						& 2.93
						& 2.68
						& 2.95
						& 2.89
						& 2.87
						& 3.14
						& 2.82 \\
						\hspace{1em} AI
						& (1.90)
						& (1.81)
						& (1.77)
						& (1.79)
						& (1.78)
						& (1.71)
						& (2.01)
						& (1.84)
						& (1.88)
						& (1.79) \\
						Fearful of AI
						& 3.17
						& 3.31
						& 3.25
						& 3.12
						& 3.21
						& 3.36
						& 3.15
						& 3.29
						& 3.15
						& 3.40 \\
						& (1.94)
						& (1.93)
						& (1.85)
						& (1.78)
						& (1.86)
						& (1.71)
						& (1.78)
						& (1.88)
						& (1.85)
						& (1.78) \\
						Willing to inter-
						& 3.46
						& 3.38
						& 3.19
						& 3.24
						& 3.30
						& 3.36
						& 3.48
						& 3.32
						& 3.36
						& 3.33 \\
						\hspace{1em} act with AI
						& (1.91)
						& (1.72)
						& (1.77)
						& (1.72)
						& (1.84)
						& (1.65)
						& (1.81)
						& (1.85)
						& (1.89)
						& (1.78) \\
						Age
						& 40.04
						& 42.40
						& 41.84
						& 42.69
						& 43.26
						& 44.61
						& 48.19
						& 44.98
						& 44.16
						& 46.38 \\
						& (15.44)
						& (14.38)
						& (16.46)
						& (17.62)
						& (16.50)
						& (17.95)
						& (16.35)
						& (17.11)
						& (17.13)
						& (17.85) \\
						Male
						& \multicolumn{1}{c}{35\%}
						& \multicolumn{1}{c}{41\%}
						& \multicolumn{1}{c}{42\%}
						& \multicolumn{1}{c}{36\%}
						& \multicolumn{1}{c}{37\%}
						& \multicolumn{1}{c}{34\%}
						& \multicolumn{1}{c}{60\%}
						& \multicolumn{1}{c}{67\%}
						& \multicolumn{1}{c}{46\%}
						& \multicolumn{1}{c}{35\%} \\
						Participants
						& \multicolumn{1}{c}{159}
						& \multicolumn{1}{c}{151}
						& \multicolumn{1}{c}{157}
						& \multicolumn{1}{c}{166}
						& \multicolumn{1}{c}{155}
						& \multicolumn{1}{c}{154}
						& \multicolumn{1}{c}{166}
						& \multicolumn{1}{c}{164}
						& \multicolumn{1}{c}{159}
						& \multicolumn{1}{c}{162} \\
						\bottomrule
					\end{tabularx}
					\footnotetext{
						\hspace{-2.15em} \textit{Note.} Mean values and standard deviations (in parentheses) of demographic characteristics. Excitement about AI, fear of AI, and willingness to interact with AI were measured on a Likert scale ranging from 0 (``not at all'') to 6 (``for sure'').
						
						Fisher's exact test was run on the variables ``male'' and ``Divert trolley'' to test for differences between treatments; analyses of variance, on the other variables. In Study~1, ethical self-assessment differs between treatments ($p = 0.033$); in Study~3, sex ($p < 0.001$). Any other differences are insignificant.
					}
				}
			\end{minipage}
		\end{turn}
	\end{center}
	
	\begin{table}[H]
		\ttabbox[\linewidth]{
			\caption{Augmented results of Studies~1--3 \label{tab:Appendix_Reg}}
		}{
			\begin{tabularx}{\linewidth}{X*{4}{D{.}{.}{5.5}}}
				\toprule
				& \multicolumn{1}{c}{Study~1}
				& \multicolumn{1}{c}{Study~2}
				& \multicolumn{2}{c}{Study~3} \\
				\cmidrule(lr){2-2} \cmidrule(lr){3-3} \cmidrule(lr){4-5}
				&
				& \multicolumn{1}{c}{Opaque}
				& \multicolumn{1}{c}{Criminal}
				& \multicolumn{1}{c}{Criminal} \\
				& \multicolumn{1}{c}{AI \&}
				& \multicolumn{1}{c}{AI \&}
				& \multicolumn{1}{c}{AI \&}
				& \multicolumn{1}{c}{AI \&} \\
				& \multicolumn{1}{c}{Human}
				& \multicolumn{1}{c}{Human}
				& \multicolumn{1}{c}{Criminal}
				& \multicolumn{1}{c}{Human} \\
				& \multicolumn{1}{c}{(1)}
				& \multicolumn{1}{c}{(2)}
				& \multicolumn{1}{c}{(3)}
				& \multicolumn{1}{c}{(4)} \\
				\midrule
				Advice to favor friend
				& -14.29^{***}
				& -14.98^{***}
				& -1.90
				& -14.81^{***} \\
				& (3.59)
				& (3.62)
				& (3.41)
				& (3.54) \\
				AI-advisor
				& 0.33
				& 2.65
				& 4.08
				& 2.93 \\
				& (3.53)
				& (3.65)
				& (3.53)
				& (3.53) \\
				AI-advisor $\times$
				& 3.89
				& 4.23
				& -14.18^{***}
				& -1.91 \\
				\hspace{1em} Advice to favor friend
				& (5.02)
				& (5.12)
				& (4.88)
				& (4.97) \\
				Moral Self-Assessment
				& 0.39^{***}
				& 0.28^{***}
				& 0.29^{***}
				& 0.26^{***} \\
				& (0.06)
				& (0.06)
				& (0.06)
				& (0.06) \\
				Divert trolley
				& -1.84
				& -6.39
				& -3.05
				& -5.67 \\
				& (3.73)
				& (4.04)
				& (4.23)
				& (3.92) \\
				Openness to AI
				& 0.96
				& 0.45
				& 0.94
				& 0.78 \\
				& (0.71)
				& (0.74)
				& (0.68)
				& (0.70) \\
				Male
				& -2.57
				& -3.51
				& -6.65^{**}
				& -2.14 \\
				& (2.62)
				& (2.73)
				& (2.58)
				& (2.62) \\
				Age
				& 0.09
				& 0.05
				& 0.04
				& 0.04 \\
				& (0.08)
				& (0.08)
				& (0.07)
				& (0.08) \\
				Intercept
				& 36.73^{***}
				& 51.70^{***}
				& 46.77^{***}
				& 52.08^{***} \\
				& (7.26)
				& (7.35)
				& (7.81)
				& (7.73) \\
				\midrule
				Observations
				& 631
				& 616
				& 649
				& 628 \\
				Groups (Scenario)
				& 3
				& 3
				& 3
				& 3 \\
				Var: Intercept (Scenario)
				& 15.23
				& 11.67
				& 35.65
				& 39.25 \\
				Var: Residual
				& 982.43
				& 998.91
				& 948.80
				& 957.78 \\
				\bottomrule
			\end{tabularx}
			\footnotetext{
				\hspace{-2.15em} \textit{Note.} Regression of moral judgment on the nature of advice and the type of advisor, with scenario as random intercept. The table reproduces the results of the regressions reported in Column~3 of Table~\ref{tab:Study1}; Column~3 of Table~\ref{tab:Study2}; Columns~3 and 5 of Table~\ref{tab:Study3}, all augmented by control variables.
				
				The participants were asked for their decision in the trolley dilemma. The variable ``Divert trolley'' takes the value 1 if the participant decided to divert the trolley and sacrifice the life of one person to save the lives of five; 0, if he or she decided not to divert the trolley. The decision to divert the trolley suggests a utilitarian mindset.
				
				Relative to the results in the text (Tables~\ref{tab:Study1}--\ref{tab:Study3}), some observations were dropped due to missing responses to the post-experimental questions.
				
				$^{**} p < 0.05$. $^{***} p < 0.01$.
			}
		}
	\end{table}
	
	Table~\ref{tab:Appendix_Reg} reports the results of similar regressions as in Tables~\ref{tab:Study1}--\ref{tab:Study3}, where we added the demographic characteristics and the answers to the post-experimental question summarized in Table~\ref{tab:PEQ} as control variables.
	
	\clearpage
	
	\section*{Pre-registration}
	
	\subsection*{Study~1}
	
	\url{https://aspredicted.org/blind.php?x=3qf7it}. %AsPredicted \#60344 (``Human and AI-based moral advisors'').
	
	\subsection*{Study~2}
	
	\url{https://aspredicted.org/blind.php?x=bf4z88}. %AsPredicted \#60654 (``AI-based moral advisors without a rationale'').
	
	\subsection*{Study~3}
	
	\url{https://aspredicted.org/blind.php?x=zs9kq4} %AsPredicted \#60970 (``AI-based advisors based on convicted criminals'').
	and
	\url{https://aspredicted.org/blind.php?x=6ib7c8}. %AsPredicted \#61083 (``Convicted criminals as moral advisors'').
	
	\clearpage
	
	\section*{Experimental Instructions}
	
	\subsection*{Study~1}
	
	Study~1 employed a $2 \times 2 \times 3$ factorial design (human advisor or algorithm; advice that it is acceptable or unacceptable to favor a friend over a stranger; business, law, or health scenario). The experiment was designed so that Screen \#3 only differed between the twelve conditions.
	
	The following screenshots include the full set of screens other than Screen \#3 and a selection of the twelve versions of Screen \#3. The versions of Screen \#3 differ only minimally, and the remaining versions can be inferred form the reprinted screenshots.
	
	\begin{figure}[h!]
		\ffigbox[\FBwidth]{
			\caption{Informed consent (Screen \#1).}
		}{
			\frame{\includegraphics[width=.75\textwidth]{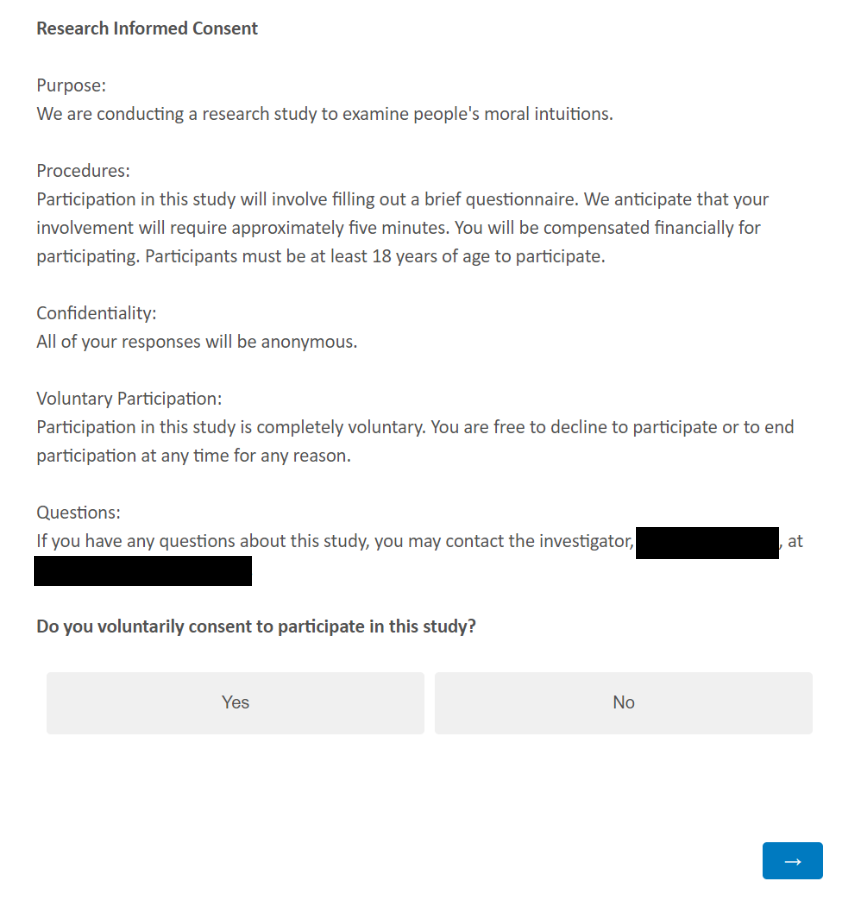}}
		}
	\end{figure}
	
	\clearpage
	
	\begin{figure}[h!]
		\ffigbox[\FBwidth]{
			\caption{General instructions (Screen \#2).}
		}{
			\frame{\includegraphics[width=.75\textwidth]{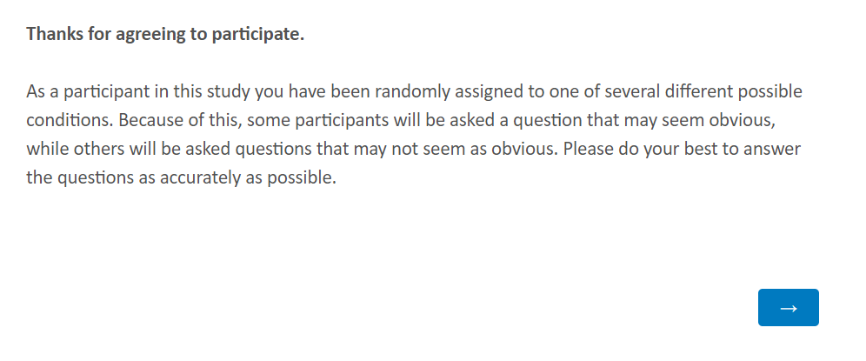}}
		}
	\end{figure}
	
	\begin{figure}[h!]
		\ffigbox[\FBwidth]{
			\caption{Decision (Screen \#3). Advice is given by a human advisor, who considers it unacceptable to favor a friend over a stranger in the business scenario. Alternatively, the advice read: ``According to the advisor, it is acceptable to hire your friend.'' The slider was set to the midpoint by default and the participant had to move it to proceed. (It could be moved back to the midpoint.) \label{fig:Human}}
		}{
			\frame{\includegraphics[width=.75\textwidth]{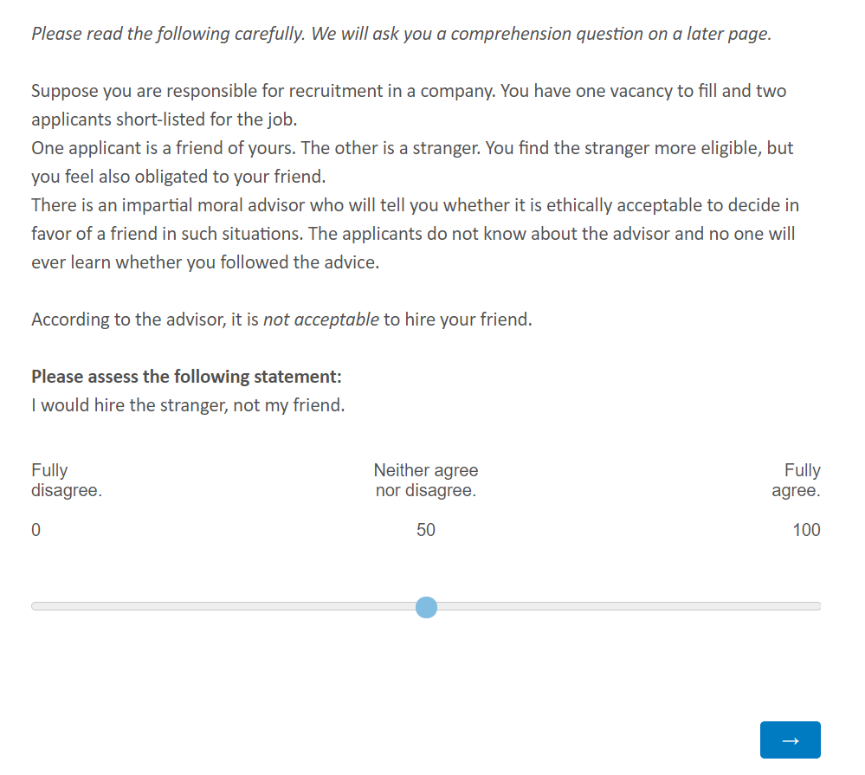}}
		}
	\end{figure}
	
	\clearpage
	
	\begin{figure}[h!]
		\ffigbox[\FBwidth]{
			\caption{Decision (Screen \#3). Advice is given by a human advisor, who considers it unacceptable to favor a friend over a stranger in the health scenario. See Figure~\ref{fig:Human} for further technicalities.}
		}{
			\frame{\includegraphics[width=.75\textwidth]{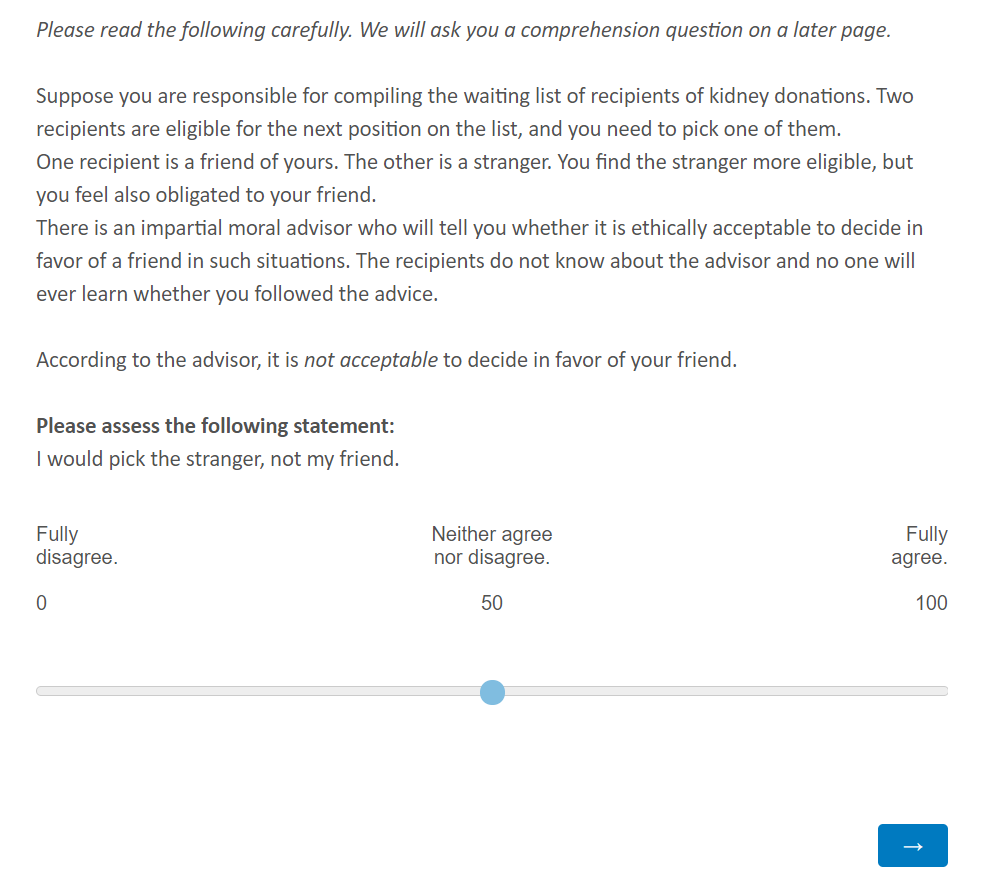}}
		}
	\end{figure}
	
	\clearpage
	
	\begin{figure}[h!]
		\ffigbox[\FBwidth]{
			\caption{Decision (Screen \#3). Advice is given by a human advisor, who considers it unacceptable to favor a friend over a stranger in the law scenario. See Figure~\ref{fig:Human} for further technicalities.}
		}{
			\frame{\includegraphics[width=.75\textwidth]{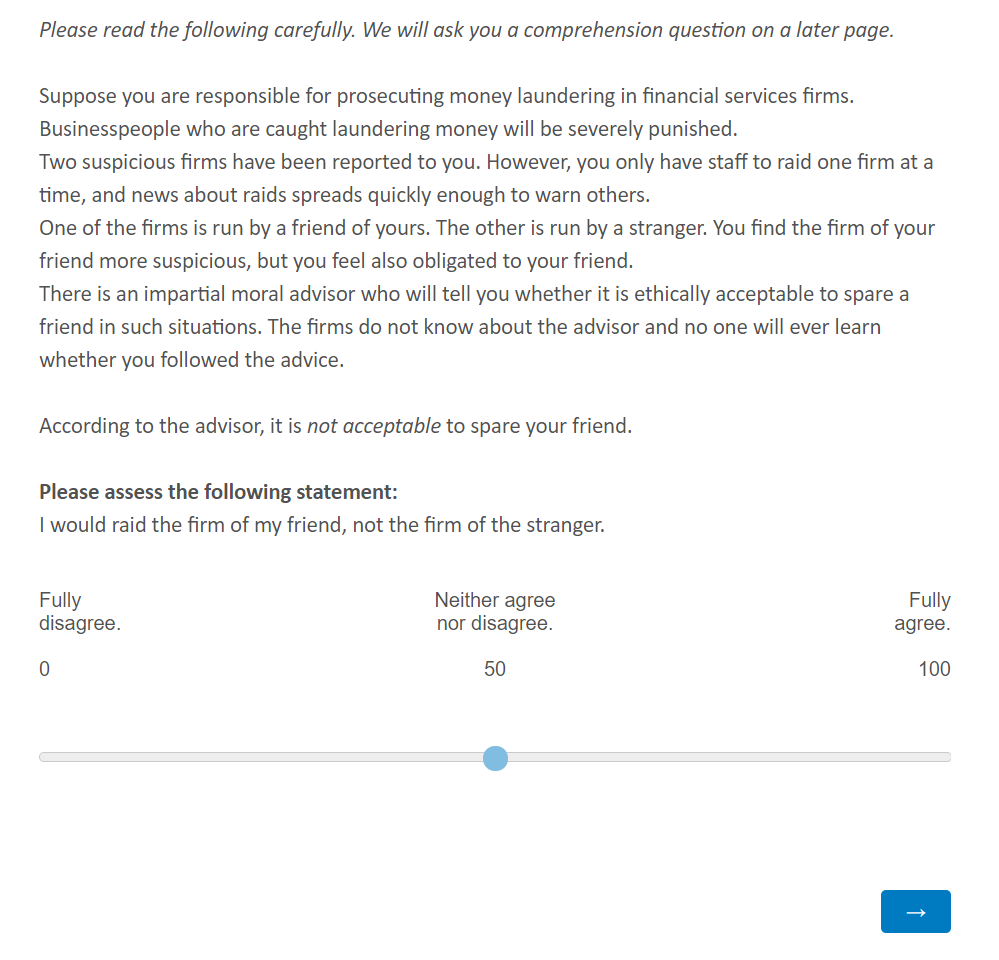}}
		}
	\end{figure}
	
	\clearpage
	
	\begin{figure}[h!]
		\ffigbox[\FBwidth]{
			\caption{Decision (Screen \#3). Advice is given by an algorithm, which considers it unacceptable to favor a friend over a stranger in the business scenario. See Figure~\ref{fig:Human} for further technicalities. \label{fig:Algo}}
		}{
			\frame{\includegraphics[width=.75\textwidth]{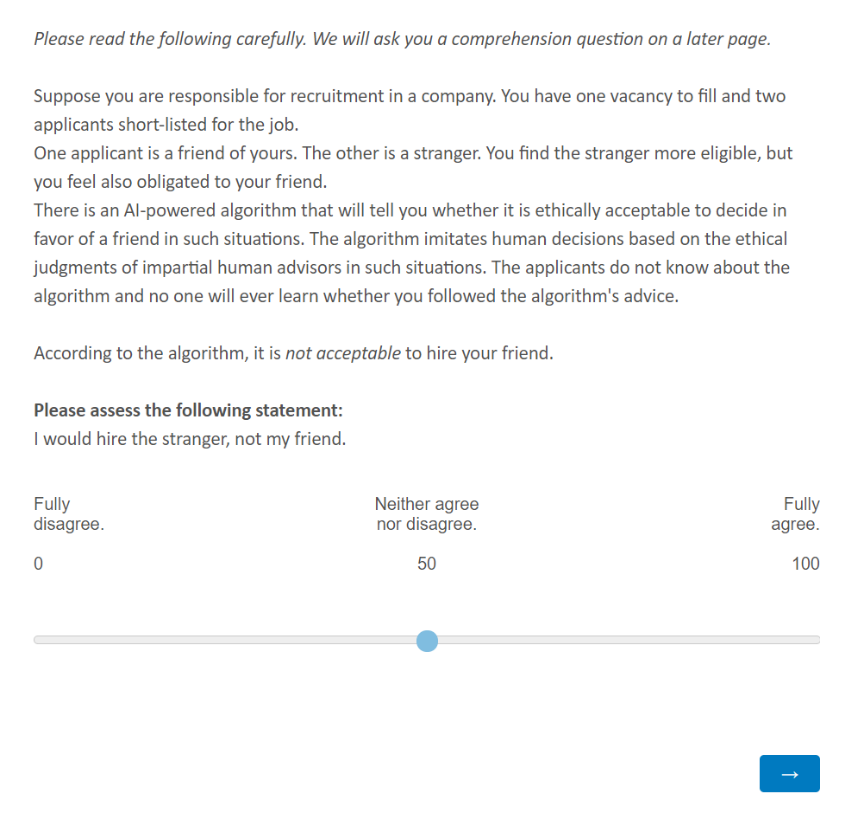}}
		}
	\end{figure}
	
	\clearpage
	
	\begin{figure}[h!]
		\ffigbox[\FBwidth]{
			\caption{Comprehension question (Screen \#4). The order of the answer options was randomized between the participants.}
		}{
			\frame{\includegraphics[width=.75\textwidth]{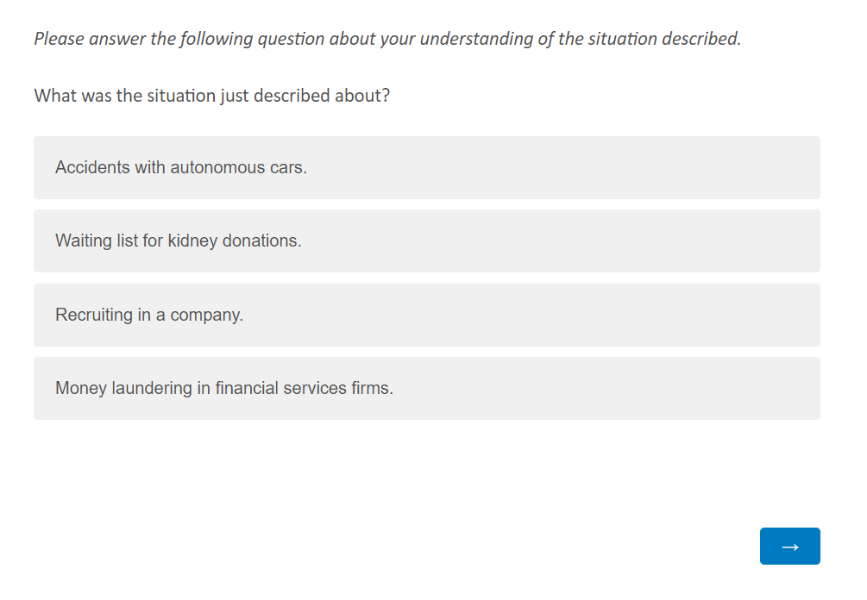}}
		}
	\end{figure}
	
	\begin{figure}[h!]
		\ffigbox[\FBwidth]{
			\caption{Influence of the on the participant (Screen \#5). The order of the answer options was randomized between the participants.}
		}{
			\frame{\includegraphics[width=.75\textwidth]{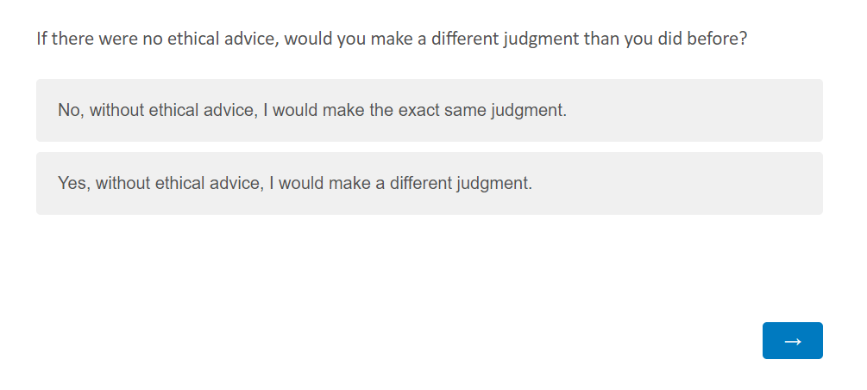}}
		}
	\end{figure}
	
	\clearpage
	
	\begin{figure}[h!]
		\ffigbox[\FBwidth]{
			\caption{Influence of advice on other participants (Screen \#6). The order of the answer options was randomized between the participants.}
		}{
			\frame{\includegraphics[width=.75\textwidth]{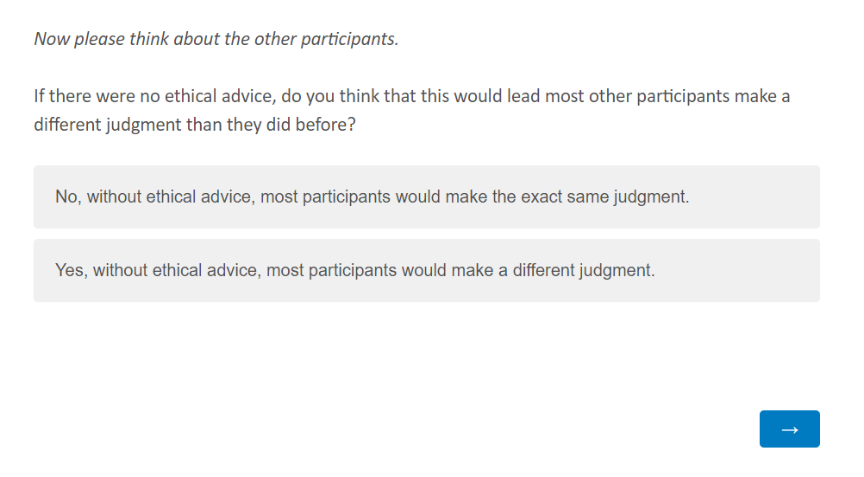}}
		}
	\end{figure}
	
	\begin{figure}[h!]
		\ffigbox[\FBwidth]{
			\caption{Ethical self-assessment (Screen \#7). The slider was set to the midpoint by default and the participant had to move it to proceed. (It could be moved back to the midpoint.)}
		}{
			\frame{\includegraphics[width=.75\textwidth]{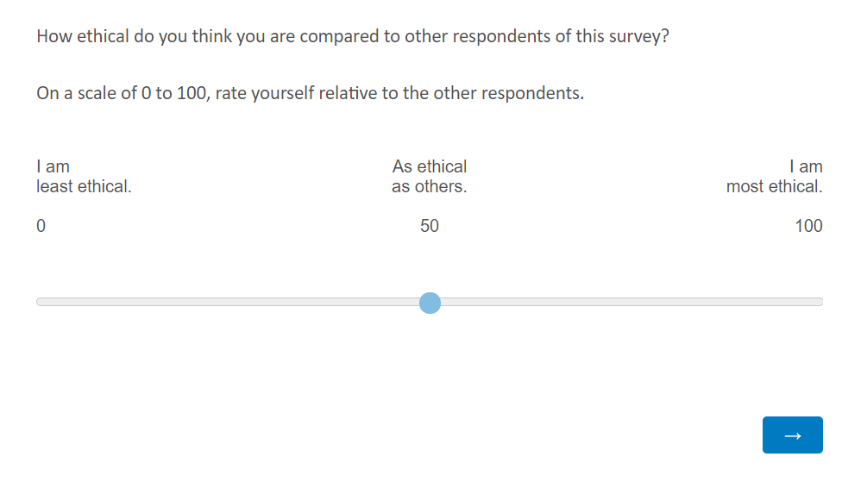}}
		}
	\end{figure}
	
	\clearpage
	
	\begin{figure}[h!]
		\ffigbox[\FBwidth]{
			\caption{Trolley dilemma to elicit the participant's ethical attitude (Screen \#8). The order of the answer options was randomized between the participants.}
		}{
			\frame{\includegraphics[width=.75\textwidth]{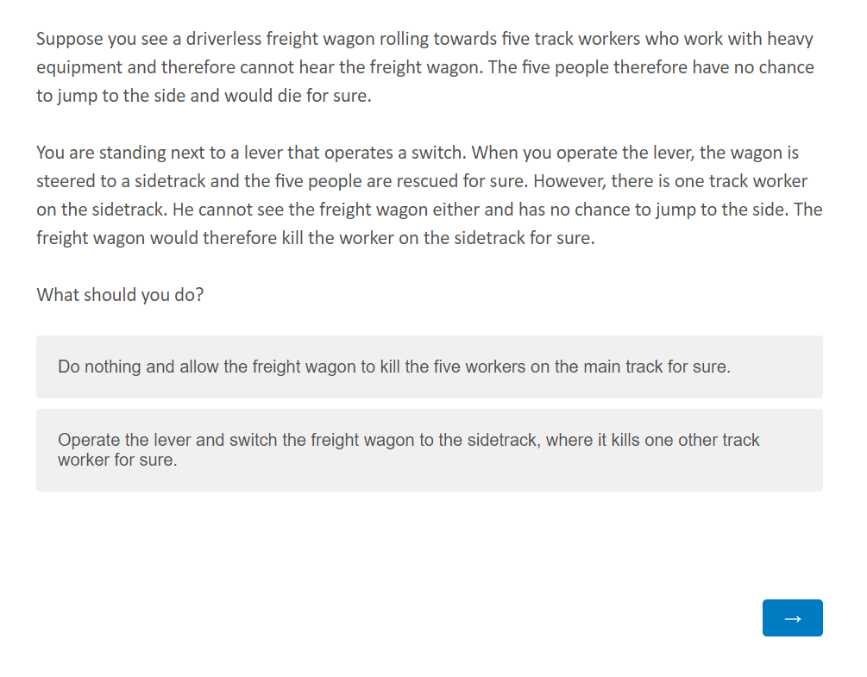}}
		}
	\end{figure}
	
	\clearpage
	
	\begin{figure}[h!]
		\ffigbox[\FBwidth]{
			\caption{Attitude toward AI (Screen \#9). The order of the questions was randomized between the participants. The answer options (shown for the last question on the screenshot) were the same for all three questions.}
		}{
			\frame{\includegraphics[width=.75\textwidth]{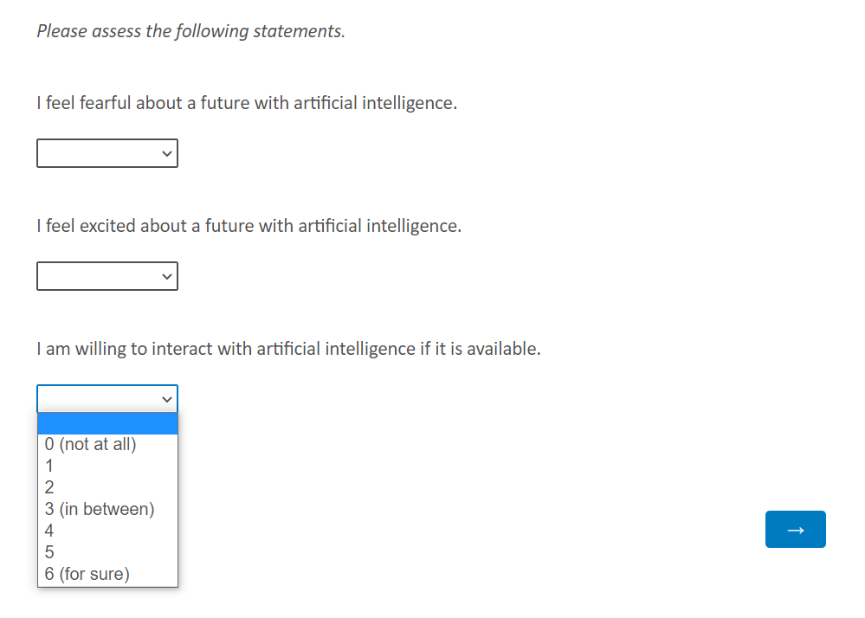}}
		}
	\end{figure}
	
	\begin{figure}[h!]
		\ffigbox[\FBwidth]{
			\caption{Demographic data (Screen \#10).}
		}{
			\frame{\includegraphics[width=.75\textwidth]{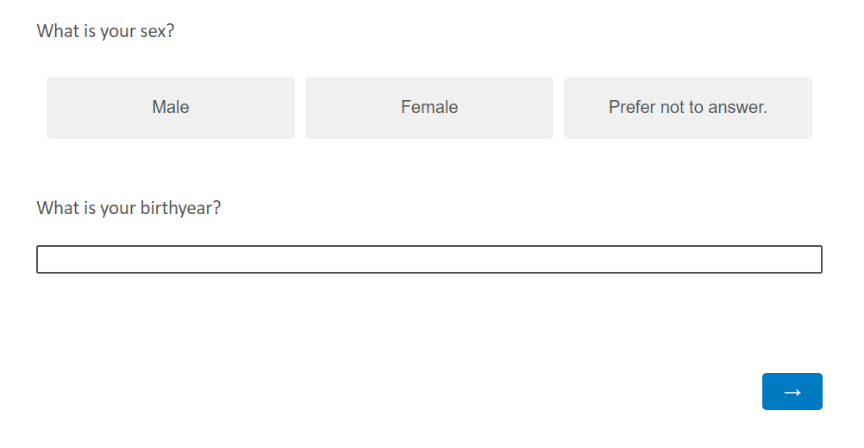}}
		}
	\end{figure}
	
	\clearpage
	
	\begin{figure}[h!]
		\ffigbox[\FBwidth]{
			\caption{End of the experiment (Screen \#11).}
		}{
			\frame{\includegraphics[width=.75\textwidth]{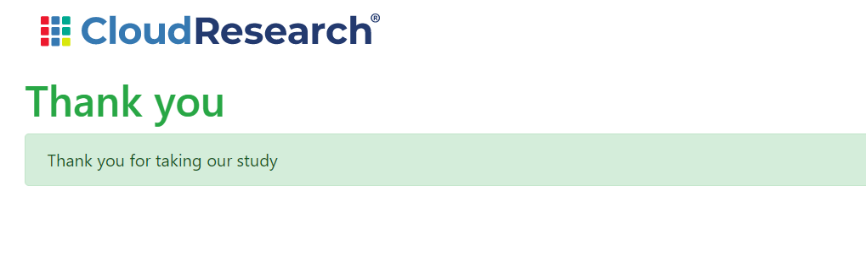}}
		}
	\end{figure}
	
	\clearpage
	
	\subsection*{Study~2}
	
	Study~2 employed a $2 \times 3$ factorial design (advice that it is acceptable or unacceptable to favor a friend over a stranger; business, law, or health scenario). The screens were identical to those of Study~1 with a human advisor except for Screen \#3.
	
	The screenshot shows Screen \#3 in the business scenario. The other five versions of the screen can be inferred like those of Study~1.
	
	\begin{figure}[h!]
		\ffigbox[\FBwidth]{
			\caption{Decision (Screen \#3). The algorithm considers it acceptable to favor a friend over a stranger in the business scenario. The text is identical to that of Study~1 (Figure~\ref{fig:Algo}) except for the information given about the algorithm.}
		}{
			\frame{\includegraphics[width=.75\textwidth]{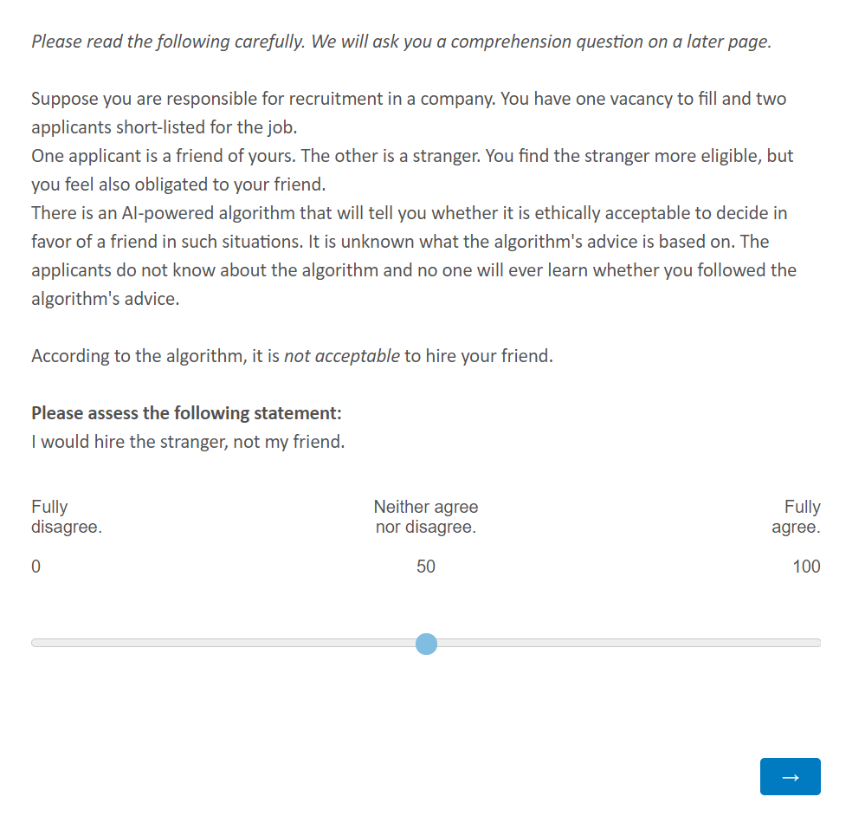}}
		}
	\end{figure}
	
	\clearpage
	
	\subsection*{Study~3}
	
	Study~3 employed the same $2 \times 2 \times 3$ factorial design as Study~1. The screens were identical to those of Study~1 except for Screen \#3.
	
	The screenshots below show Screen \#3 in the business scenario. The other ten versions can be inferred in the same way as in Study~1.
	
	\begin{figure}[h!]
		\ffigbox[\FBwidth]{
			\caption{Decision (Screen \#3). Advice is given by a convicted criminal, who considers it unacceptable to favor a friend over a stranger in the business scenario. See Figure~\ref{fig:Human} for further technicalities.}
		}{
			\frame{\includegraphics[width=.75\textwidth]{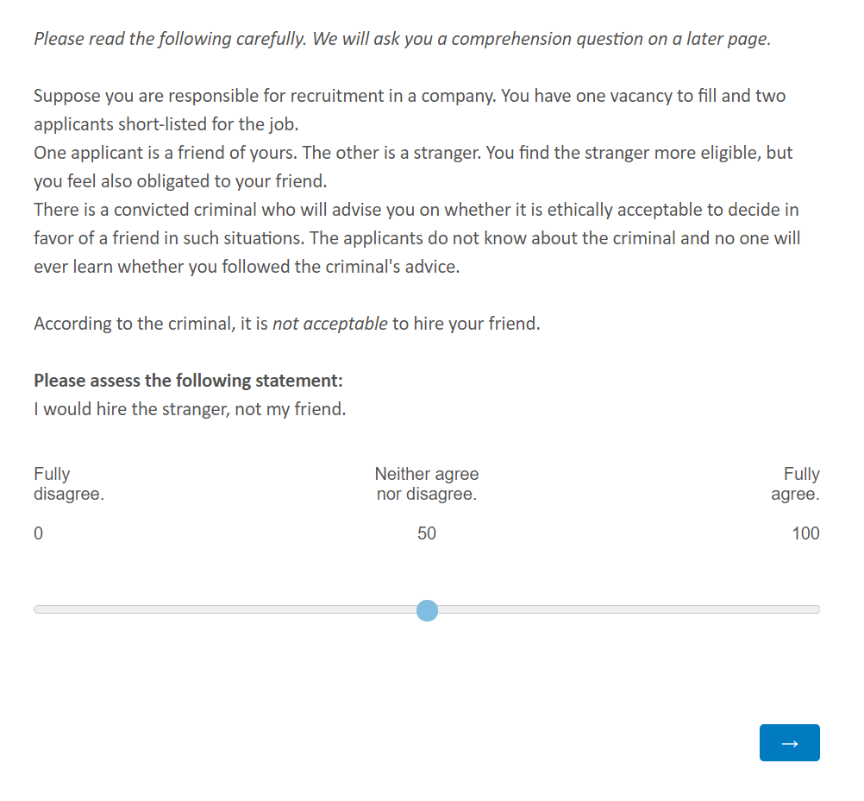}}
		}
	\end{figure}
	
	\clearpage
	
	\begin{figure}[h!]
		\ffigbox[\FBwidth]{
			\caption{Decision (Screen \#3). Advice is given by an algorithm, which is modeled on convicted criminals and which considers it unacceptable to favor a friend over a stranger in the business scenario. See Figure~\ref{fig:Human} for further technicalities.}
		}{
			\frame{\includegraphics[width=.75\textwidth]{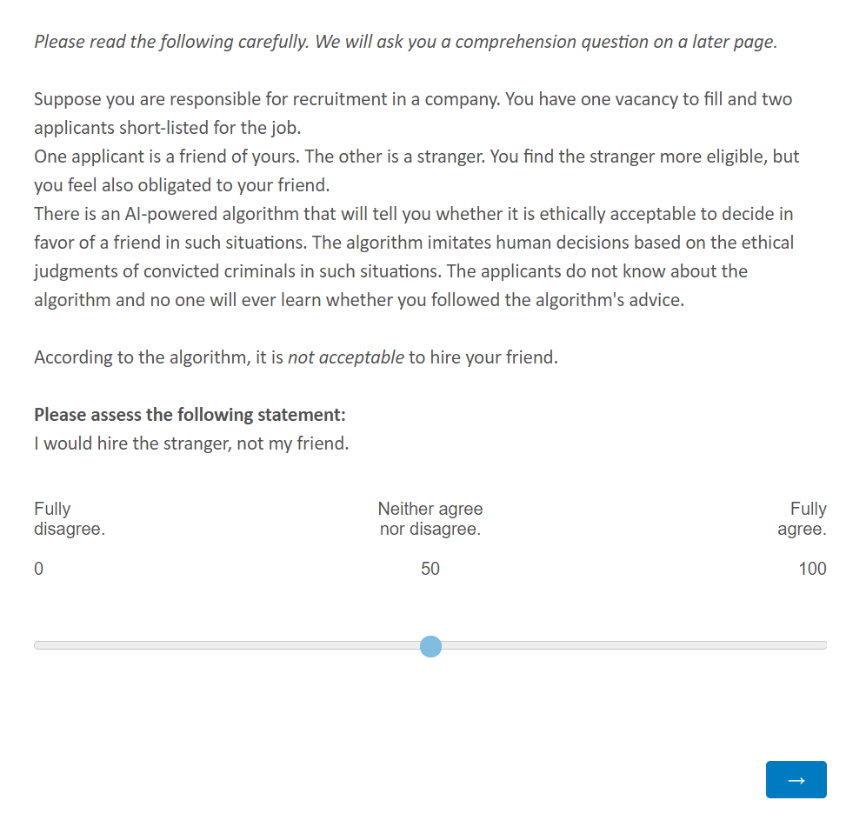}}
		}
	\end{figure}
\end{document}